\documentclass{revtex4}
\input epsf
\usepackage{epsfig}
\newcommand {\h}{\frac{1}{2}}

\newcommand {\eq}{\begin{equation}}
\newcommand {\qe}{\end{equation}}
\newcommand {\cen}[1]{\begin{center} #1 \end{center}}
\newcommand {\bfr}{{\bf r}}

\newcommand {\ot}{\frac{1}{3}}
\newcommand {\twt}{\frac{2}{3}}

\newcommand {\di}{<\frac{1}{d}>}
\newcommand {\fmm}{fm$^{-1}$}
\newcommand {\nucp}{Nucl. Phys. }

\begin{document}


\title {Isospin breaking from diquark clustering}

\author{W. R. Gibbs}

\affiliation{\large New Mexico State University, Las Cruces,
NM 88003}

\author{Jean-Pierre Dedonder}

\affiliation{\large
Sorbonne universit\'es, Universit\'e Pierre et Marie Curie,\\
Sorbonne Paris Cit\'e, Universit\'e Paris-Diderot et
IN2P3-CNRS, UMR
7585,\\
Laboratoire de Physique Nucl\'eaire et Hautes Energies,\\   
4, Place Jussieu, 75252 Paris cedex 05, France}
\date{\today}

\begin{abstract}   

\begin{description}\item[Background]

Although SU(2) isospin symmetry is generally assumed in the basic theory 
of the strong interaction, a number of significant violations have been 
observed in scattering and bound states of nucleons. Many of these 
violations can be attributed to the electromagnetic interaction but the 
question of how much of the violation is due to it remains open.

\item[Purpose]

To establish the connection between diquark clustering
in the two-nucleon system and isospin breaking from the
Coulomb interaction between the members of diquark pairs.

\item[Method] 

A schematic model based on clustering of quarks in the interior of the 
confinement region of the two-nucleon system is introduced and evaluated. 
In this model the Coulomb interaction is the source of all isospin 
breaking. It draws on a picture of the quark density based on the 
diquark-quark model of hadron structure which has been investigated by a 
number of groups.

\item[Results] 

The model produces three isospin breaking potentials connecting 
the unbroken value of the low-energy scattering amplitude to 
those of the $pp$, $nn$ and $np$ singlet channels. A simple test of the 
potentials in the 3-nucleon energy difference problem yields 
results in agreement with the known binding energy difference. 

\item[Conclusion] 

The illustrative model suggests that the breaking seen in the low-energy 
NN interaction may be understood in terms of the Coulomb force between 
members of diquark clusters. It allows the prediction of the charge 
symmetry breaking interaction and the $nn$ scattering length from the well 
measured $np$ singlet scattering length. Values of the $nn$ scattering 
length around $-18$ fm are favored. Since the model is based on the quark 
picture, it can be easily extended, in the SU(3) limit, to calculate 
isospin breaking in the strange sector in the corresponding channels. A 
natural consequence of isospin breaking from diquark clustering 
is that the breaking in the strange sector, as measured by the separation 
energy difference between $_{\Lambda}^4$H and $_{\Lambda}^4$He,  is 
several times larger than that seen in the comparison of three-nucleon
mirror nuclei as observed experimentally. 
\end{description}

\end{abstract}
\maketitle

{\large
\section{Introduction}

In spite of the very significant progress that has been made in the study 
of Quantum Chromodynamics (QCD), the understanding of the low-energy 
nucleon-nucleon (NN)  interaction in terms of parton degrees of freedom 
remains elusive. The short-range part of the potential has been calculated 
with parton degrees of freedom in a number 
of cases, often with one gluon exchange (which also involves two quark 
exchange) in first order perturbation theory. Such theories tend to give 
no (or little) attraction. Since the two-pion exchange produces a strong 
intermediate-range attraction \cite{parispot} it might be thought that 
this mechanism alone would suffice. An attempt \cite{parisq} to marry the 
short range quark contribution \cite{takeuchi,okaqm,thirdmodel} to the 
well known one- and two-pion exchange did not succeed. While empirical 
evidence of two-pion exchange exists \cite{2pi} it appears that it does 
not provide all of the intermediate range potential strength needed.

When the confinement volumes of two nucleons begin to overlap, 
direct quark-quark interactions can occur through 
non-color-singlet exchanges producing quark clustering among 
quarks originating in different nucleons. Underlying this picture 
is the notion that the NN interaction involves sharing of quarks 
between the two incident particles when they overlap, 
thus contributing to the NN potential. Even if this clustering 
occurs in an isospin symmetric manner (which we assume), the 
different charges of the quarks will produce energy differences 
which will break the SU(2) symmetry. It is this source of isospin 
breaking that we treat in this paper. Since the Coulomb 
interaction is hidden in the correlations, we refer to the effect 
as ``crypto-Coulomb'' (CC). We treat only the valence quarks even
though the sea quarks would be expected to participate as well in a 
more realistic calculation.

In a diquark-quark model of the nucleon the third (unpaired) 
quark in each nucleon can combine in the overlapping and enlarged 
confinement region thus adding to the intermediate range 
attraction. Models in which quark-quark interactions have been 
included in a single baryon have been moderately 
successful \cite{bloch,carlson1,carlson2} in predicting the mass spectrum.

Here we focus on a more modest goal than producing a full 
NN potential; instead, we look at the {\em result} 
of this type of interaction and study the Coulombic effect of 
quark clustering in the valence quark density and compare the result 
with experimentally observed isospin breaking to infer the 
parameters describing the quark density.

Not having available the strong potential which would be the result 
of such a strong-interaction model, we use a phenomenological 
potential of the Reid soft-core type \cite{reid} which has a 
short-range repulsion, medium range attraction and one-pion-exchange 
(OPE) tail to provide the basic SU(2) invariant interaction. This 
potential is determined by fitting $pp$ scattering data below 30 MeV 
and then used to calculate the $nn$ and $np$ effective range 
parameters and compare them with data. The entire process uses a 
unique quark density (same for $nn$, $pp$ and singlet $np$ pairs) 
with the isospin breaking being produced solely by the charge 
differences between the valence quarks themselves.

Section \ref{diquarkrev} reviews the evidence for diquark clustering 
in the interior of hadrons along with a discussion of the expected 
role of diquark clustering in low-energy NN scattering. 
In Section \ref{review} a brief summary of the data on isospin 
breaking in the NN interaction and the status of the theory are 
reviewed. In Section \ref{crypto} a simple model is introduced which 
is based on the na\"ive quark model but satisfies many of the conditions 
just discussed. In section \ref{extract} the method used for 
extracting the strong phase shifts from the data is given.  Section 
\ref{results} gives the results of the model. An extension of these 
ideas to the strange sector ($\Lambda$-N) is presented in Section 
\ref{strange}. Section \ref{discuss} gives a short discussion with 
conclusions.

\section{Diquark models\label{diquarkrev}}

While diquark models have been around for some time, there
has been an increase in interest recently due to new experimental
results. In this section we review briefly this history.

The model of nucleon as a diquark (D) quark (q) pair is as old as the 
idea of quarks themselves. When Gell-mann introduced the name 
``quark'' he discussed the possible existence of the diquark at the 
same time \cite{gellmann}. Ida and Kobayashi \cite{ida}, and 
Lichtenberg and Tassie \cite{tassie} were the first to consider the 
diquark-quark model. Ono \cite{ono} and Lichtenberg et al. 
\cite{lichtenberg,lichtenberg1} studied the mass spectrum which would 
be generated by such a model. Abbott et al. \cite{abbott} saw 
evidence for a spin-one diquark in the need for an explanation of the 
large ratio of longitudinal to transverse photon absorption. Gunion 
\cite{gunion} saw the need for a diquark in the form of the structure 
functions of the nucleon. Later Mineo et al.~\cite{mineo2} calculated 
the structure functions with a diquark model.

The existence of diquark correlations in the nucleon is claimed to be 
supported by a large body of experimental evidence (see ref. 
\cite{rmp} for a summary of the situation in 1993). Lattice gauge 
theory has provided some theoretical support 
\cite{alexandrou,hess,babich,forcrand} although the size of the 
diquark extracted is often larger than that assumed by typical models,
but not always~\cite{engel}. 

Much remains to be known about the detailed properties of these 
correlations. There seem to be indications for both a spin-zero and a 
spin-one form. The spatial extent of the correlations also varies with 
the specific model treated. A typical result from the lattice is that 
the spatial size of the axial vector diquark is larger than that of 
the scalar diquark. Weiss et al. \cite{weiss}, in a Nambu-Jona-Lasinio 
model, found that a larger axial vector diquark was necessary in order 
to have the square charge radius of the neutron be negative. See also 
Ishii et al. \cite{ishii} for a similar model.

One diquark model of the nucleon which is particularly successful 
is the one due to Keiner \cite{keiner}. While many diquark-quark 
calculations do not respect isospin invariance, Keiner presented 
a diquark model for the nucleon which gives an excellent fit to 
the electromagnetic properties of the nucleon and does respect 
SU(2) symmetry. In order to conserve isospin he included both 
scalar and axial vector diquarks of the same mass and size 
(0.24 fm).

More modern treatments \cite{guo1,weng} deal with heavy baryons. 
Clo\"et, Bentz and Thomas \cite{cloet} use diquarks to calculate 
clustering effects on the nucleon elastic form factors.
Santopinto and Ferretti \cite{santopinto} treat strange and
non-strange baryons. Another model which leads to quark-quark correlations is 
the Flux-tube model \cite{carlson1,carlson2}.

More recently, a diquark-antidiquark structure has been 
studied \cite{maiani1,maiani2,maiani3,maiani4,brinkstancu2,
bigi,dubnicka,terasaki1,terasaki2,yyang,kleiv,piccinini,
ebert3,ebert1,ebert2,ebert4,jmr,padmanath} in the form of 
[$Qq$][$\overline{Q}\overline{q}$] for recently observed particles 
by the BaBar \cite{aubert}, Belle \cite{choi}, and LHCb \cite{aaij} 
collaborations. 

Of perhaps more relevance for NN scattering are the nearly 
degenerate light scalar mesons [$f_0(980)$\ and $a_0(980)$] which 
have long been a problem to understand in a $q\bar{q}$ model. They 
have often been viewed as meson-meson molecular states but more 
recently the interest has increased to study them in the 
diquark-antidiquark picture 
\cite{weinisgur1,maiani5,deng,ebert6,brinkstancu1,weinisgur2, 
zhang}. In summary, it seems that the idea of diquark clustering 
in the internal structure of hadrons, light or heavy, is 
widespread and has many supporters.

Diquark models would seem to have taken a blow when it became evident 
that the valence quarks carry only a fraction of the spin of the 
proton. However, Myhrer and Thomas \cite{myhrer1} and Thomas 
\cite{thomasprl} pointed out that relativistic effects and the pion 
cloud could explain this apparent lack. Shortly thereafter, 
Clo\"et and Miller published \cite{cloetmiller} a diquark-quark model 
of the nucleon in which the axial vector diquark plays an essential 
role by contributing negatively to the spin sum thus reducing the 
apparent fraction of the spin carried by the quarks. In their fit to 
the electromagnetic form factors they found the axial vector diquark 
to have a {\it smaller} mass than the scalar diquark, contrary to 
most other models, a point that remains to be understood.

Although we know of no current theory which takes into account the 
clustering of pairs of quarks into diquarks for the NN interaction
in free space, there have been treatments of the nucleonic interaction
in nuclear matter in Landau-Migdal models \cite{horikawa,bentztony}.
There is also a study of the transition of nuclear matter to
quark matter where a condensate of diquarks is needed to obtain the
phase change \cite{bentzhori}. The rearrangement of quarks into
quark pairs has also been treated in the meson-meson interaction 
\cite{lenz}.  

It is hard to imagine that quark clustering will not play some role 
in the short-to-mid-range $NN$ interaction. When the two confinement 
volumes begin to overlap (when they touch) the quarks will be free to 
move about in the aggregated volume so that the two (initially) 
unpaired quarks can combine to form a diquark thus lowering the total 
mass of the two-nucleon system thereby creating an attractive 
potential. This picture is especially favorable at very low energies 
where the overlap of the two systems exists for a long time allowing 
the quarks ample opportunity to rearrange themselves.

The construction of such a theory is well beyond the scope of this paper. 
Instead, we look at the connection between the clustering which might be 
generated by such a theory and isospin breaking observed in the NN system 
in the singlet scattering lengths.

\section{Overview of isospin breaking in the NN 
interaction\label{review}}

In 1974 it was suggested \cite{sauer2} that it might be possible 
to learn about the short-range NN interaction from 
the study of isospin breaking in  
this interaction. It is the 
purpose of this paper to present a schematic model realizing this 
idea using as the only source of breaking the Coulomb 
interaction between quark pairs.

SU(2) symmetry (isospin) is one of the fundamental building blocks of 
QCD. It is known that this symmetry is broken by the Coulomb 
interaction but it is not known to what extent it is broken in the 
strong interaction. 

One of the most obvious manifestations of isospin breaking 
is the difference in the nucleon masses. It is generally 
believed that the neutron-proton ($np$) mass difference is due to 
the up-down quark mass differences 
\cite{davies,bicker,science,gandl}. While the $ud$ quark 
mass difference is of the same order as the interior 
Coulomb effects we will treat, it will not enter into our 
considerations for the following reason: while the 
different singlet nucleon pairs will indeed have different 
total masses, they will not depend on the distance between 
the centers of mass of the nucleons so that there is no 
localized potential to use for the calculation of the 
scattering or binding of nucleons. The result is constant 
energy shifts which are expressed as the difference in 
nucleon masses. The nucleon masses will affect the kinetic 
energy contribution to the bound states and the phase space 
of the scattering states leading to a small breaking that 
can be treated in the hadronic calculations needed to 
remove the conventional Coulomb effects.

Other examples of isospin breaking exist in nuclear physics 
and it is important to know if they come from additional 
fundamental breaking or if they can be explained by the 
electromagnetic interaction. A special case of the breaking 
is a failure of charge symmetry (CSB) wherein the $nn$ 
interaction is different from the ``Coulomb corrected'' 
$pp$ interaction. A classification of the types of isospin 
breaking was given by Ref. \cite{henleymiller}. One review 
of the breaking as of 1990 was given in Ref. \cite{review}, 
another in 2006 \cite{reviewa} and again in 2009 
\cite{gard}.

By making the exchange neutron $ \Leftrightarrow $ proton 
in a nuclear system, i.e., by comparing the energies of 
mirror nuclei, a measure of the CSB can be obtained. In 
order to carry out this program the ordinary Coulomb 
interaction must be taken into account. A great deal of 
work has been done in this direction.

\subsection{The Okamoto-Nolen-Schiffer anomaly}

Discussions of the CSB problem often start with the 
Okamoto-Nolen-Schiffer anomaly \cite{okamoto,ns} which is the 
observation that the addition of the conventional Coulomb 
force between protons does not give enough energy difference 
between members of a nuclear isospin doublet. By the 
conventional Coulomb force we mean a $e^2/r$ potential at 
large distances with a continuation for smaller $r$ 
representing some sort of nucleon charge density, often taken 
as a uniformly charged sphere. We denote this potential by 
$V_C(r)$. The same discrepancy is observed in the binding 
energy difference between $^3$He and $^3$H (see next 
section).

\begin{figure}[htb]
\epsfig{file=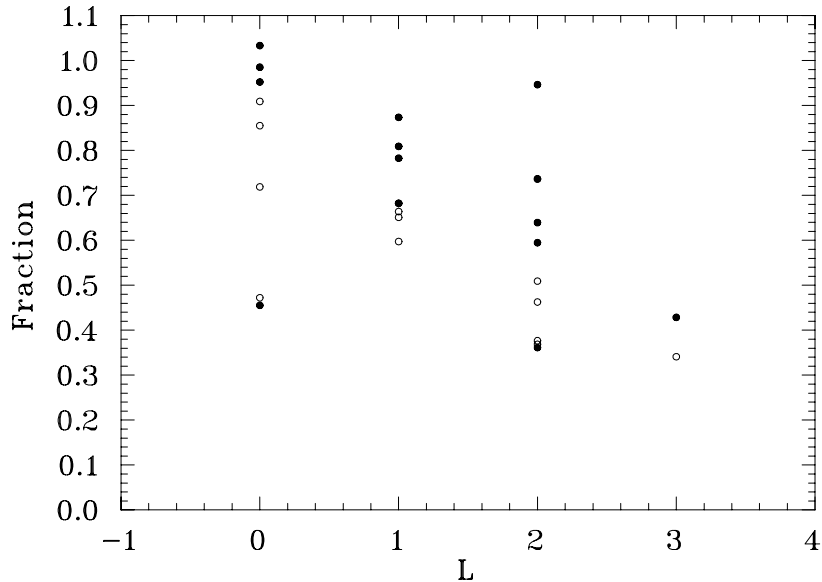,angle=90,height=3.in}
\caption{The dots represent the ratio of the calculation of 
Blunden and Iqbal \cite{blunden} based on the $\rho-\omega$ 
and $\pi-\eta$ charge symmetry breaking potentials of \cite{sid} 
to the CSB energies of Sato \cite{sato} as a function of the
angular momentum of the shell. The solid dots correspond 
to the SKII results \cite{vb}, the open circles to DME 
calculations \cite{nv}. }
\label{blunden}\end{figure}

 In an attempt to explain this discrepancy by a difference in the 
fundamental interaction the exchange of heavy mesons (primarily 
the mixing of the $\rho$ and $\omega$ mesons) \cite{sid} has been 
proposed \cite{blunden}. These mesonic CSB potentials \cite{sid} 
give the right order of magnitude \cite{blunden} but they appear 
to fail quantitatively for the higher angular momentum shells as 
can be seen in Fig. \ref{blunden} where the fraction of the 
needed additional energy provided is shown. The discrepancy in 
the prediction seems to increase for the larger systems.

Sato \cite{sato}, who did the calculation of the Coulomb energy 
from the nuclear wave functions from which the inadequacy of the 
pure ordinary Coulomb force is obtained, attempted to fit the 
missing energies with an arbitrary charge-symmetry-breaking 
potential constrained only to have a range, less than or equal 
to, one pion exchange but did not find a fit. The calculations of 
Blunden and Iqbal \cite{blunden} were done with shell-model wave 
functions and hence did not have a short-range repulsive 
interaction excluding the NN wave function from 
short relative distances. Since the mesonic CSB potentials 
\cite{sid} have a very short range (essentially the same as the 
one-boson-exchange often used to provide the source of the 
strongly repulsive short-range potential), one might question the 
validity of this procedure since it might be expected to overestimate 
the effect.

\subsection {Three-body comparisons}

While the mirror nuclei comparison offers several useful 
features, in particular the possibility to change the range over 
which the breaking interaction is being sampled by varying the 
atomic number, there is a complication in that a reliable 
structure calculation is needed. As seen from Fig. \ref{blunden} 
considerable variation is present from the difference in the two 
types of shell-model calculations \cite{vb,nv} considered.

The comparison of the binding energy of $^3$He and $^3$H offers a 
way of examining charge symmetry breaking in which the 
uncertainties due to nuclear structure are largely absent. An 
additional advantage is that the short-range repulsion of the 
unbroken NN interaction can be taken into account by 
the use of ``realistic'' potentials. The conventional Coulomb 
force can be included in Fadeev calculations to any desired 
accuracy.

In the three-nucleon system it is found that the conventional Coulomb 
potential supplies about 85\% of the difference in binding energy 
(648 keV of the 764 observed) and the remaining 15\% can be 
reliably attributed to CSB beyond what has been put in to the 
calculation so far. Wu, Ishikawa and Sasakawa \cite{wu} 
considered a number of small effects (including magnetic 
interactions and the kinetic energy effect of the $np$ 
mass difference) and concluded that about 46 keV of the 116 keV 
needed could be supplied by the sum of these small effects 
leaving about 70 keV to be attributed to further CSB 
interactions. Friar, Gibson and Payne \cite{friar1} found a very
similar result.

They estimate that a potential which represents the shift of 
scattering lengths in the singlet S state of -17.5 fm to -18.6 fm 
(see section \ref{scatt} for a discussion of the source of these 
determinations)  will also give the right value for the binding 
energy difference. The $\rho$-$\omega$ CSB potential \cite{sid} 
provides a large fraction of what is needed in their calculation. 
For the breaking of charge independence they assumed a 
phenomenological Woods-Saxon potential with a radius of 0.5 fm, a 
diffuseness of 0.2 fm, and a 6 MeV depth. The calculation is 
complicated by the fact that the absolute binding energy of $^3$H 
is not given correctly so that an extrapolation must be made.

\subsection{NN Scattering states\label{scatt}}

Scattering states are sensitive to different aspects of the breaking 
potentials than the bound states so, in principle at least, additional 
information can be obtained by studying scattering cross sections.

There is a wide range of values the $^1$S$_0$ NN 
scattering length that have been obtained experimentally (Table 
\ref{aNN}), besides the $np$ singlet scattering length first 
measured long ago in $np$ scattering from ortho- and 
para-hydrogen \cite{squires,stewart,koester}. It was found to be 
definitely negative leading to the conclusion that there was no 
lightly bound state of two neutrons. The study of the differences 
in these values has been the subject of many papers.

The $pp$ scattering length, $-17.3 \pm 0.4$ fm, \cite{review} is 
obtained from the removal of the Coulomb interaction in $pp$ 
scattering \cite{ppdat}, a procedure which depends on the strong 
interaction used and the assumption of the form for the short 
range Coulomb potential. If this number is taken to represent the 
unbroken value then it is being assumed that all of the breaking 
in the proton-proton ($pp$) scattering is due to the conventional 
Coulomb interaction.

\begin{table}
\begin{tabular}{|c|c|c|} \hline
{$a_{NN}$\ (fm)} &{\rm Reaction}&{\rm Reference}\\
\hline
-23.748 $\pm$ 0.009&{\rm np\ scattering\ from\ ortho-\ and\ 
para-hydrogen}&\cite{squires}-\cite{koester}\\
\hline
-17.3 $\pm$ 0.4&{\rm Coulomb\ corrected\ pp\ scattering}&
\cite{review,reviewa,ppdat}\\
\hline
-18.5 $\pm$ 0.3 &{\rm nn\ final\ state\ interaction\ in}\ $\pi^- d \to nn\gamma$\
{\rm and nd\ break-up}&
\cite{gabioud1}-\cite{chen}, \cite{ggs}\ {\rm and}\ \cite{gtprl}-\cite{shirato}\\
\hline
-16.1$\sim$ -16.3&{\rm nn\ final\ state\ interaction\ in\ nd\ break-up} 
&\cite{review,reviewa}\cite{huhnprl,huhn} \\
\hline
\end{tabular}
\caption{Experimental determinations of the singlet 
NN scattering length }.
\label{aNN}
\end{table}

Neutron-neutron $nn$ scattering lengths with values around $-16.1 \sim 
-16.3$ fm obtained from the final state interaction in $nd$ breakup 
were considered to be the recommended values at one time but now 
the value, $-18.5 \pm 0.3 $ fm, mostly from the $\pi^-d\rightarrow 
nn\gamma$ reaction is favored by the majority of physicists working 
in this field. This preference may be because the values from $nd$ 
breakup have a large dispersion and those from radiative pion 
absorption experiments are consistent \cite{howelltalk} but also
because the values around -18~fm are in the correct direction
relative to $a_{pp}$ to explain the binding-energy difference
in the three-nucleon system.

The value extracted from the $np$ interaction is far from the 
group of the others. Since this value indicates a system closer to 
being bound it would seem to require an additional attractive 
potential (relative to the others) to move the value in that 
direction.

The breaking expressed by the difference of the value of the $np$
scattering length compared with the average of the $nn$ and $pp$
values is usually considered as a separate effect (charge
independence breaking, CIB) from the charge symmetry breaking
(the difference between $a_{nn}$ and $a_{pp}$). Since only
neutral pions can be exchanged in $nn$ and $pp$ scattering, the
fact that the charged pion mass is different from that of the
neutral pion does not enter into CSB. However, it does contribute
to the CIB. The replacement form for one-pion exchange when the
masses are different was worked out some time ago
\cite{henleypi,sidmike} and has been claimed to provide much
\cite{cheung} or even all \cite{ericmiller} of the breaking
needed. Cheung and Machleidt \cite{cheung} point out that Ericson
and Miller \cite{ericmiller} used an approximation which, when
corrected, reduces their value to something comparable with other
determinations. All of these corrections were calculated without
form factors or with form factors with large regulating masses.
We find that this correction is sensitive to the value of the
cut-off parameter as will be discussed later in the text.

The role of two pion exchange in CIB has been the subject of 
several studies, resulting in a wide range of values for the 
contribution to the difference from the average $nn$ and $pp$ 
values of the singlet $np$ value. Some values are: 0.65 fm 
\cite{friarcibpi}, 0 \cite{sidmike}, 0.16 fm \cite{limach}, 0.18 
fm \cite{cheung}, and 0.88 fm \cite{ericmiller}.

The pattern of these scattering lengths, which should all be the 
same under SU(2) symmetry, would seem to require three 
isospin-breaking interactions: one to connect the unbroken value 
to the $pp$ scattering, one to connect it to the $nn$ scattering 
and a third to connect it to the singlet $np$ scattering. One 
might suppose that the isospin-pure value would lie somewhere near 
the centroid of these values which would mean that the correction 
for the $pn$ case would be represented by a attractive potential 
while the $nn$ scattering would be produced by a repulsive 
potential. The $ pp$ potential would have to be more repulsive 
than the $nn$ case to give the right sign for the mirror nuclei 
difference observed. The considerations in the following sections 
naturally lead to three potentials which fulfill these conditions.

\subsection{A challenge to the meson-exchange theory of CSB}

This picture seems to have changed when Goldman, Henderson and 
Thomas \cite{terry} calculated the effect of a form factor in the 
derivation of the CSB $\rho$-$\omega$ potential in a quark picture 
and found that the resulting potential is much smaller than 
previously believed. While one might question the quark model 
used, their paper was shortly followed by one by Piekarowicz and 
Williams \cite{jorge1} using hadronic ($N\overline{N}$) 
intermediate states which found qualitatively the same result and 
one by Hatsuda et al. \cite{henley} which found a similar result 
using QCD sum rules. Soon after followed several more papers which 
argued that these calculations were only examples of a general 
rule that the contribution to CSB from $\rho$-$\omega$ mixing 
should be negligible 
\cite{krein,maltman,oconnell1,oconnell2,mitchell} (but see 
\cite{coonrebut,cohen}). Thus, it appears that a replacement 
theory is needed.

One such alternative theory was the treatment of the effect of 
nucleon mass differences in two-pion exchange. The major 
contributor is the crossed diagram for the exchange of two charged 
pions with nucleon or $\Delta$ intermediate states. The difference 
of the internal nucleon masses, compared with the external masses, 
changes sign for $nn$ compared with $pp$ scattering.

Coon and Niskanen \cite{coonnisk} found that the difference in 
scattering lengths could be understood with this effect in spite 
of an earlier calculation which had found the effect to be 3-5 
times smaller. Li and Machleidt \cite{limach} treated the same 
diagrams and also found that they could understand the symmetry 
breaking. However, the two calculations are very different. Coon 
and Niskanen \cite{coonnisk} find that the diagram with two 
nucleons in the intermediate state dominates while Li and 
Machleidt \cite{limach} find that it is the diagram with one 
$\Delta$ which is, by far, the largest. Machleidt and M\"utter 
\cite{machmut} investigated the possibility of choosing between 
the heavy meson exchange contribution and the two-pion exchange by 
studying the partial waves with $\ell>0$.

The paper by Coon and Niskanen \cite{coonnisk} contains a summary 
of the calculations of the two-pion exchange (before the work of 
Li and Machleidt) and concludes that ``This cannot be considered a 
satisfactory theoretical situation.'' These calculations were done 
with regulating masses of the pion form factor of the order of 1 
GeV. Coon and Niskanen \cite{coonnisk} show that the scattering 
length difference is sensitive to the value chosen. For the values 
of this mass that we use (see section \ref{pimass}) the 
contribution of these diagrams would be very small.

\section{Coulomb effects due to clustering \label{crypto}}

\subsection{Intra-nucleon Coulomb} 

First we consider the Coulomb energy which arises from the 
assembly of several charges. The total energy contained in such a 
charge cloud depends very much on the details of how the charges 
are clustered. An example is provided by the nucleons themselves.

The proton has two valence quarks ($u$) with charge +2/3 and one 
($d$) with charge --1/3. If they are equidistant from each other 
with an average inverse distance of $<\frac{1}{d}>$ then the total 
coulomb energy of the system is

\eq e^2\left[ \frac{2}{3} \times
 \frac{2}{3}- \frac{1}{3}\times \frac{2}{3}-  \frac{1}{3}\times 
\frac{2}{3} \right]<\frac{1}{d}>=0 \qe

For the neutron we have

\eq e^2\left[ \frac{1}{3} \times  \frac{1}{3}- \frac{1}{3} \times
 \frac{2}{3} - \frac{1}{3}\times \frac{2}{3} ,\right]<\frac{1}{d}>= - 
\frac{1}{3}\ e^2<\frac{1}{d}>. \qe

This result is perhaps less surprising when one considers that an 
alternate decomposition, $n\rightarrow p+\pi^-$ (the pion cloud), 
would give a negative Coulomb energy as well. There is no reason, 
however, to believe that all of the distances between quarks are 
the same, indeed one popular model has two of the quarks combined 
in a diquark which has a small size. Of course, this effect still 
has the wrong sign to explain the $np$ mass difference; 
invoking fundamental quark mass differences appears to remain 
necessary.

This form of interior Coulomb effect has been known for a long
time and forms the basis for predictions of the electromagnetic
mass differences of Baryons and Mesons (see, for example, 
\cite{rubin,agal,lichtmass} and \cite{gandl} and references
therein.

\subsection{Coulomb effects interior to the confinement range of two 
nucleons}

When two nucleons are at small values of the center-of-mass 
separation the six quarks intermixed give more possibilities for 
correlations among them leading to additional Coulomb-energy 
effects. We restrict this freedom by correlating two of the quarks 
in each nucleon to represent a diquark (D). The view that we have 
is that when the two confinement regions touch, and begin to 
overlap, the partons can move around in the combined volumes. The 
initially unpaired quarks will be able to pair up and form a 
diquark correlation. Of course this will happen only a fraction of 
the time so that some probability will be associated with the 
event. It is plausible that the motion of these components will be 
governed by the interaction with the gluons within the expanded 
confinement region so that the pairing will be largely independent 
of the initial state of the nucleons and the breaking will be 
state independent. Of course, within a nucleus, states which 
differ in having a different probability for the overlap of the 
nucleons would show a state dependence.

To obtain an estimate of the scale of Coulomb energies for a 
single diquark, consider using an exponential shape for the 
density of the relative motion of the two quarks forming the 
diquark. In this case we have \eq 
E_C=\frac{1.44\sqrt{3}n}{<r^2>^{\h}9}\approx \frac{0.277n}{<r^2>^{\h}} {\rm 
MeV},\qe where $<r^2>^{\h}$ is the r.m.s. radius of the 
correlation density in fm (from 0.25 to 1 fm from different 
models) and $n=+4, +1$ and $-2$ for $uu, dd$, and $ud$ pairs.

Any model for Coulomb energy due to quark clustering in the 
two-nucleon system will have (at least) three physical 
parameters:

1) The overall size of the system, expressed here as the diameter 
of a six quark bag. This parameter will control the size of the 
Coulomb effect between the uncorrelated quarks and the range of 
the resulting interaction. Representing the combined system as a 
sphere is a crude approximation, since, as the interaction 
proceeds, one would expect a deformation of the individual 
confinement regions as well as a possible ``necking'' of the 
combined system.

2) The range of the correlations. This (along with the electric 
charges) controls the Coulomb energy contained in each correlated 
quark pair.

3) The probability of forming a quark-quark correlation. This can 
be expressed as the probability of the two-quark state being found 
in a given configuration, given that it is known to be found in a 
known volume.

The following model contains these three pieces of physics in a 
simplified form. While the illustrative model to be presented 
shortly does not assume point diquarks, it is useful in order to 
get a qualitative picture to consider this example. In this case 
one has four clusters of charge to deal with. The result depends 
on which two quarks are taken to form the diquark.

A common assumption (which we shall follow) is to take the unlike 
quarks in the nucleon to form the diquark (one $u$ and one $d$ 
giving a charge +1/3 object). In this case one has for the CC 
interaction between two neutrons

$$ C_{nn}(r)=e^2\left(\frac{1}{3}\times 
\frac{1}{3}\di_{DD}+\ot\times \ot\di_{qq}-(\ot\times 
\ot+\ot\times \ot)\di_{Dq}\right) 
$$ 
\eq 
=\frac{e^2}{9}\left(\di_{DD}+\di_{qq}-2\di_{Dq}\right) 
\label{eq1}\qe 
where $\di_{DD}$ is the average inverse distance between the 
diquarks in the two nucleons, $\di_{qq}$ is the average inverse 
distance between the non-associated quarks in each nucleon (we 
will assume shortly that this pair of quarks also forms a diquark 
but for the moment it is general) and $\di_{Dq}$ is the average 
inverse distance between the odd quark in one nucleon and the 
diquark in the other. There are of course other Coulomb energies 
from quarks in the interior of each nucleon but these do not 
depend on the distance between the centers of mass of the two 
nucleons (or so we will assume) and does not contribute to the
crypto-Coulomb potential.

For a neutron and a proton we have
$$
C_{np}(r)=e^2\left(\ot\times \ot\di_{DD}-\twt\times 
\ot\di_{qq}+(\ot\times 
\twt-\ot\times \ot)\di_{Dq}\right).
$$
\eq
=\frac{e^2}{9}\left(\di_{DD}-2\di_{qq}+\di_{Dq}\right)
\label{eq2}\qe
and 
$$
C_{pp}(r)=e^2\left(\ot\times \ot\di_{DD}+\twt\times 
\twt\di_{qq}+(\ot\times 
\twt  +\twt\times \ot)\di_{Dq}\right)
$$
\eq
=\frac{e^2}{9}\left(\di_{DD}+4\di_{qq}+4\di_{Dq}\right)
\label{eq3}\qe
for the two proton case. Clearly the "ordinary Coulomb potential
between two protons is included in this last expression as well. 

The dependence on $r$, the center-of-mass separation of the 
nucleons, comes from the dependence of the average inverse of the 
distance between quarks on it.

If one assumes that the odd quarks (in different nucleons) become 
correlated so as to have a large inverse distance, one can obtain 
the negative potential needed for the $np$ scattering 
length. A moderately strong effect is also due to interaction in 
the $DD$ system. However, it is the same for all nucleon pairs so 
does not make a contribution to the breaking. Neglecting 
$<\frac{1}{d}>_{Dq}$ we see that $C_{np}(r)$ is attractive 
(making the scattering length more negative than the unbroken 
case) while both $C_{nn}(r)$ and $C_{pp}(r)$ are repulsive 
(making their scattering lengths less negative). Of course it is 
well known that the difference in the neutral and charged pion 
masses makes a significant contribution to the CIB. This effect 
is discussed in section \ref{pimass}.

\subsection{Schematic model\label{model}}

For our purposes we do not need to know the wave function of the 
quarks since the density is enough to calculate the Coulomb 
energies. We consider a (modest) range of diquark sizes and vary 
the strength of the correlation to fit the $np$ singlet 
scattering length. We make use of Ockham's razor in the 
calculation by not discussing any feature that is not directly 
relevant.

We neglect any isospin breaking in the strong interaction; all 
violations come from these crypto-Coulombian potential energies. 
Thus, there is a single density which is the same for any singlet 
NN pair, only the charges of the quarks are varied according to 
the particular nucleon pair. 

\begin{table}
$$\begin{array}{|cccc|}
\hline
{\rm quark\ pair}&pn&nn&pp\\
\hline
1-4&+4&+4&+4\\
1-5&-2&-2&-2\\
2-4&-2&-2&-2\\
2-5&+1&+1&+1\\
\hline
1-6&+4&-2&+4\\
2-6&-2&+1&-2\\
3-4&-2&-2&+4\\
3-5&+1&+1&-2\\
\hline
3-6&-2&+1&+4\\
\hline
\end{array}$$
\caption{ Multiple of the factor 1/9 for the charges on individual
quarks. Quarks 1 and 2 are members of a diquark in one nucleon and
4 and 5 form a diquark in the other while 3 and 6 are unpaired in
the separated nucleons but are correlated when the nucleons overlap.
The model expectation value for the inverse distances is 
the same within each group (between the lines) allowing the addition
of the coefficients, resulting in the same values as in Eqs.
 \ref{eq1}, \ref{eq2}, and \ref{eq3}. 
}\label{factors} 
\end{table}

One can look ahead to anticipate the results of the model 
qualitatively before making the calculation. Since the 
interaction of the members of the intranucleon diquarks are all 
the same and the interactions of the odd quark in one nucleon 
with the members of the diquark in the other nucleon are also all 
the same, the interactions can be grouped as shown in Table 
\ref{factors}. This grouping leads us back to Eqs.~\ref{eq1}, 
\ref{eq2} and \ref{eq3} with the interpretation of the average 
inverse distances corresponding to the groups in Table 
\ref{factors}. Thus, we need only calculate the three inverse 
distances, which are the same for all nucleon pairs, as a 
function of $r$ and use the combinations given in the table to 
get an estimate of the CC contribution for each case. From this 
the relative contributions can be calculated immediately. We 
first note that the $DD$ contribution (first term) is the same 
for all nucleon pairs so leads to a shift away from the isospin 
conserved value which is the same for all nucleon pairs and hence 
is not visible in the differences among nucleon pairs that are 
observed experimentally.

The Coulomb interaction between the odd quark in one nucleon with 
the diquark in the other nucleon (third term in the equations, 
group two in the Table) is relatively small since no correlation 
between them is assumed. Thus the interaction between the two odd 
quarks will dominate. We can then get an estimate of the relative 
size of the various breaking potentials from the coefficients. 
The $np$ potential will have an attractive potential with weight 
two and the $nn$ potential will have a repulsive potential with 
weight one giving the breaking between $nn$ and $np$ a weight of 
three. The $pp$ potential will have a weight of four so that the 
breaking between the $pp$ and $np$ systems will have a weight of 
six or twice the difference between the $pp$ and $nn$ potentials. 
Of course the pion mass difference also contributes to the CIB. 
The factor of four in the $pp$ case is less clean than assumed above 
because the comparison is usually made with the ``Coulomb 
corrected'' value of the $pp$ scattering length and hence depends 
on the form of the Coulomb potential used at short range.

Since both scalar and axial vector versions of diquarks have been 
proposed, the nucleon wave function might consist of a linear 
combination of the two. We don't need to know that, however. We 
only need to know that there is a quark-quark correlation between 
the odd quarks in the two-nucleon system and it needs to be the 
same for all nucleon pairs to have the amplitude due to the 
strong interaction be invariant under SU(2). The correlation 
which is assumed to form between the two odd quarks need not have 
the same properties as the correlation already present in the 
individual nucleons, although we assume here that it does 
to reduce the number of parameters. The correlation between the 
odd quarks is assumed to always be the same whether they are like 
or unlike quarks (which might seem to indicate that the 
correlation is spin-one). Without that assumption a large 
breaking of isospin symmetry would be present in the strong 
interaction.

We now present a representative model calculation based on the 
na\"ive quark model considering only the valence quarks. While 
this model is too crude to believe in any detail, one may hope 
that the basic physics is contained in this simple form. There 
are no dynamics in the model; it is represented by a static quark 
density. The diquark in this case is embodied by a correlation 
between two of the quarks in each nucleon and a correlation 
representing the formation of a diquark consisting of the 
combination of the remaining unpaired quarks in the initial 
nucleons. We are considering the scattering of very low energy 
nucleons so the time available for this formation is very long 
and the combined two-nucleon system is considered as a static 
object.
  
The density is expressed as a product of correlation functions. 
The quarks are considered as distinguishable and are numbered 1 
through 6. Quarks 1, 2, and 3 are in one nucleon and 4, 5 and 6 
are in the other. Confinement is represented by a limit of the 
distance between pairs of quarks. More complicated (and more 
realistic?) conditions could easily be included. The confinement 
and correlation functions are the same for all NN 
pairs and are as follows:

pairs 1-2 and 4-5 are confined and correlated,

pairs 1-3, 2-3, 4-6, 5-6 are confined but not correlated,

pair 3-6 (the odd quarks) are correlated but not confined,

all other quark pairs have no constraints.

The correlation is represented by a Jastrow style factor 

\eq 
F_c(d)=1+\alpha e^{-\beta d} 
\qe 

where $d$ is the distance between two quarks and  
$\alpha$ is defined in terms of $\beta$ by 
\eq \alpha=\frac{1}{2} \ p\beta^3
\qe
thus adjusting the normalization so as to compensate for the 
range of the correlation. It is the parameter $p$ which is 
varied. With this choice of normalization, $p$ remains of the
order of 1-5 fm$^3$ as we will see later.

As discussed above we assume that the correlation between ``odd'' 
(unpaired) quarks in different nucleons is the same although they form 
a $ud$ pair in the $np$ case and a $uu$ or $dd$ pair in the cases of 
like nucleons.

The confinement factor is
\eq
G(d)=\frac{1}{1+e^{(d-c)/a}}
\qe
where $a$ is chosen to be very small ($10^{-4}$ fm).  The length $c$ 
represents the diameter of a spherical bag containing the 6 quarks.

Now that the model nucleon density has been defined, the
breaking potential can be calculated by taking the expectation
value of the Coulomb energy over the density for fixed chosen
values of the center-of-mass distance (the usual adiabatic
approximation) which we have already remarked can be reduced
to the calculation of the average of three inverse distances.
The calculation needed is an 18 dimensional (6 particles in 3
dimensions) integral over a highly correlated integrand with
the normalization to be determined. It is likely that several
judicial changes in integration variables would permit a
reduction in the number of dimensions (perhaps even
considerably) but we treat the integral as it stands with the
Metropolis algorithm, which could still be applied in a more
general case (say, with a truly 3-body confinement formula).
This method also readily permits specifying the distance
between the centers of mass.}

\begin{figure}[htb]
\epsfig{file=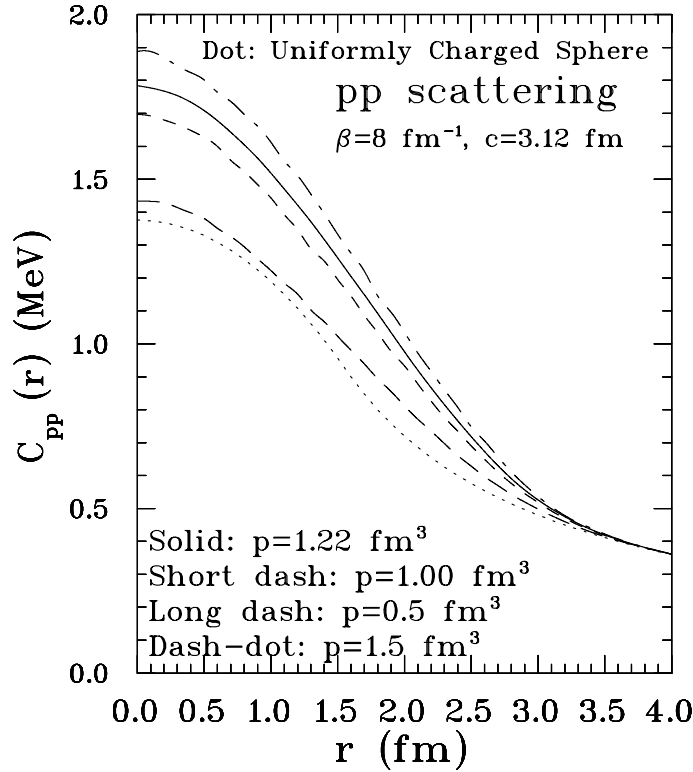,height=4in}
\caption{Variation with parameters of the model for the 
full $pp$ Coulomb potential.\label{potpps}} 
\end{figure}
\begin{figure}[htb]
\epsfig{file=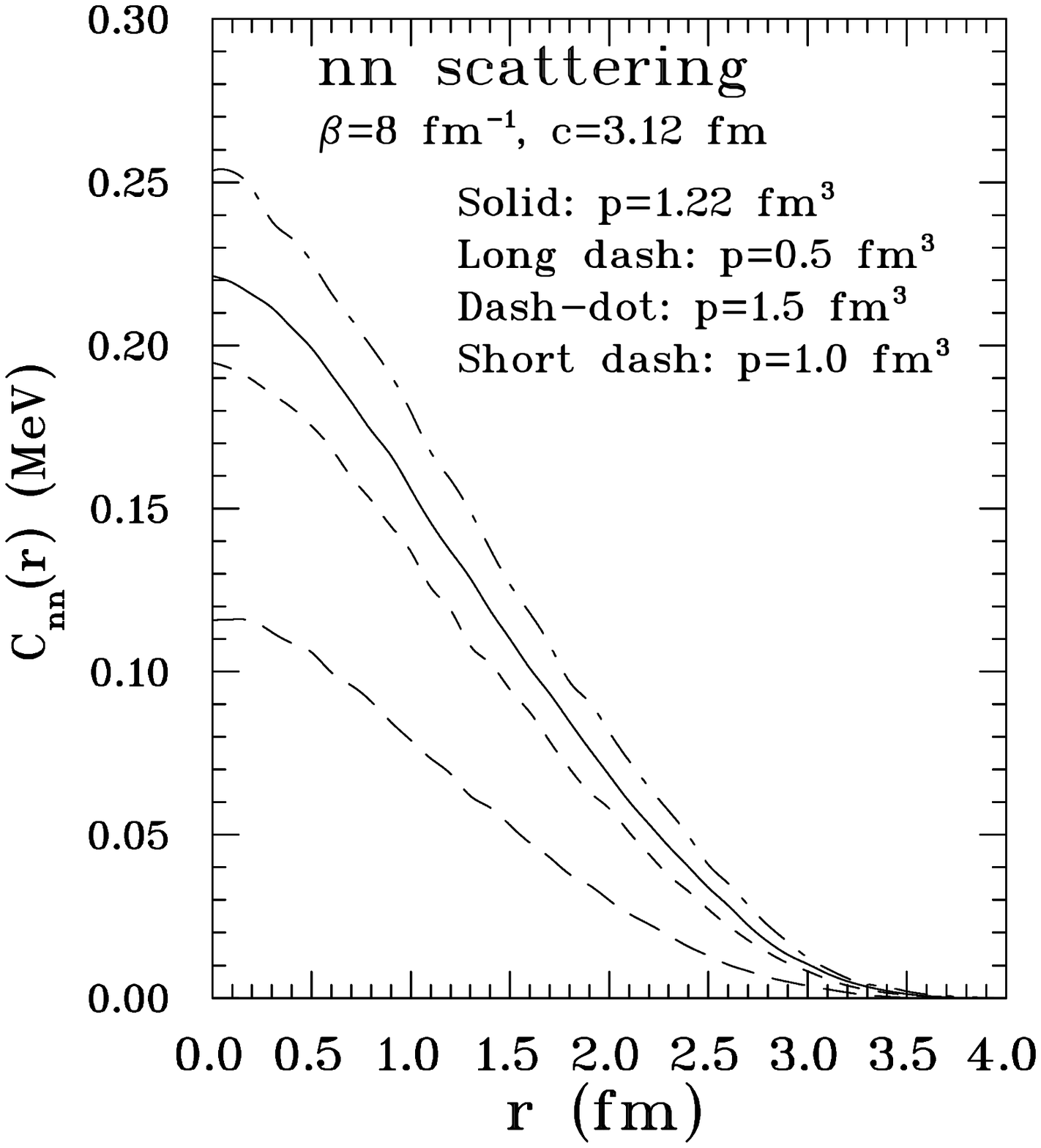,height=4in}
\caption{Variation with parameters of the model for the 
$nn$ crypto-Coulomb potential.\label{potnns}} 
\end{figure}
\begin{figure}[htb]
\epsfig{file=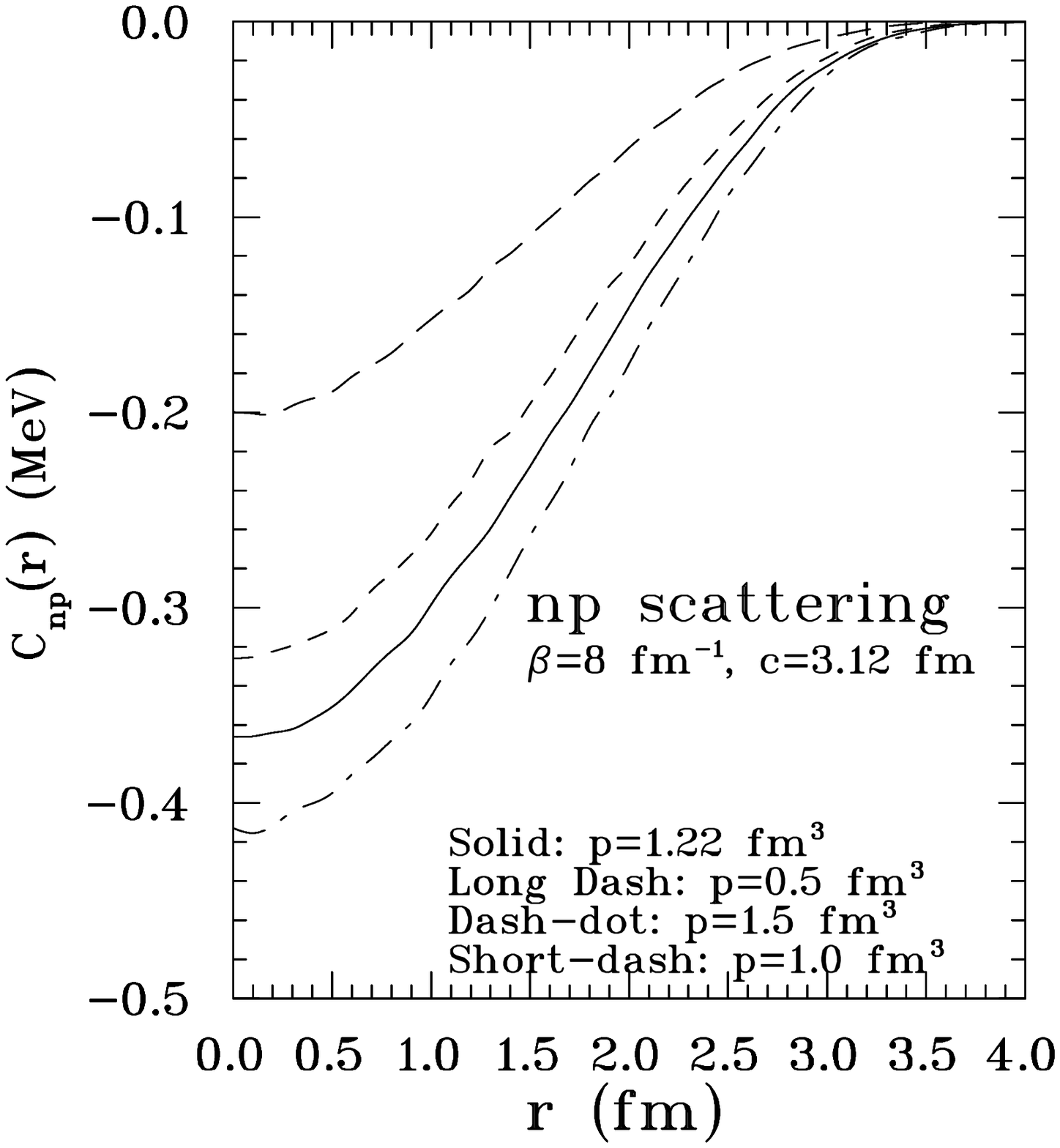,height=4in}
\caption{Variation with parameters of the model for the 
$np$ crypto-Coulomb potential.\label{potpns}} 
\end{figure}

The Metropolis algorithm creates a list of configurations 
(walkers) which consists of sets of the 18 coordinates giving the 
possible positions of all of the particles. The frequency of the 
appearance of a given configuration (or similar configurations) 
gives the probability for that configuration. In order to fix the 
distance between the centers of mass it is sufficient to restrict 
the possible walkers to those with the desired separation. To do 
this, the two nucleons are created with their centers of 
mass located at the origin at each Metropolis step. Then, 
formally, one nucleon is translated from the origin by the 
center-of-mass separation specified for any given calculation. In 
practice this translation can be combined with the evaluation of 
the distances needed for the calculation of the walker 
coordinates. The distances between quarks in the same nucleon are 
given by $\bfr_i-\bfr_j$ ($i$, $j$ in the same nucleon) while the 
distances between quarks in different nucleons are given by 
$\bfr-\bfr_i+\bfr_j $ where $\bfr$ is the distance between the 
centers of mass of the two nucleons, $i$ is 1, 2, or 3 : $j$ is 
4, 5, or 6 and the conditions $\bfr_1+\bfr_2+\bfr_3=0$\ and 
$\bfr_4+\bfr_5+\bfr_6=0$\ are maintained throughout the 
calculation. The quark mass differences that would be needed to 
explain the $np$ mass difference are not included in 
the present calculation. Thus there is no SU(2) breaking in the 
strong interaction in the model; all of the breaking comes from 
the Coulomb interaction.

The calculation was carried out with a Metropolis step size of
0.03 fm with 4000 walkers and 200,000 steps. The calculation
for one value of the relative distance between centers of mass
requires 25 minutes on a 2.5 GH CPU. With a 4 core processor
the calculation of the CC potential to 4.0 fm in steps of 0.1
fm takes less than 5 hours 

Figures \ref{potpps}, \ref{potnns} and \ref{potpns} show some 
examples of the potentials obtained with the model which result 
from the Coulomb interaction between the correlated 
quark pairs. The $pp$ potential in Fig.~\ref{potpps} includes the full Coulomb
potential (CC+``ordinary'').

\subsection{Completing the potential}

In order to actually compare with the data, full potentials must 
be available. It is assumed that there must be an isospin pure 
strong-interaction potential which is the same for all three 
singlet-nucleon pairs. The parameters of this potential are 
determined by fitting the accurate $pp$ data including the full 
Coulomb (both ordinary and CC) potential for the $pp$ case.

An isospin invariant potential given (in MeV) by a form similar
to the one assumed by Reid \cite{reid},
\eq
V_0(r)=\frac{-C_1\ e^{-\mu_1x}+C_2\ e^{-\mu_2x}}{\mu_0r}
-10.47\ \frac{(e^{-\mu_0r}-e^{-M r})}{\mu_0r}
\label{reidpot}\qe
is employed for the $nn$ and $pp$ cases. The values $\mu_1=4 
$ \fmm \ and $\mu_2=7$ \fmm \ were used by Reid \cite{reid} 
and we have kept the same ranges except where specified 
otherwise. Here $x=0.6945\ r$, $\mu_0 = 0.6840$ is the 
neutral pion mass in \fmm, while $M$ is the mass governing 
the square of the pion-nucleon off-shell form factor (see 
discussion below). For the $np$ interaction where charged 
pion exchange contributes, the third term is replaced by 
\cite{henleypi,sidmike}

\eq
10.47\left[\frac{(e^{-\mu_0r}-e^{-M r})}{\mu_0r}
-2\left(\frac{\mu_+}{\mu_0}\right)^2\frac{(e^{-\mu_+r}-e^{-M 
r})}{\mu_0r}\right]\label{ope1}
\qe
where $\mu_+$ is the mass of the charged pion. 

\subsection{Contribution of the difference of pion masses to
the breaking of charge independence\label{pimass}}

The difference in mass between the charged and neutral 
pions has long been recognized \cite{henleypi,sidmike} as 
an important contributor to the breaking of charge 
independence. Coon and Scadron \cite{sidmike} found that 
mass differences in the exchange of heavy mesons and 
two-pion exchange generated only small breaking effects. 
We find that the degree of breaking from the pion mass 
difference depends rather sensitively on the off-shell 
form factor. The form factor lowers the strength of the 
one-pion-exchange (OPE) potential at short distances and 
since we require that the total potential must be such 
that the $pp$ scattering data are correctly fit, 
the fraction of OPE potential becomes less and the 
breaking due to the pion mass difference along with it.

The correction for the finite size of the pion-nucleon 
system is often expressed as a vertex function. If this 
vertex function is denoted by $v(q)$ then the modified 
one-pion-exchange potential will be given by

\eq
V_{\pi}(r)=\frac{f^2}{4\pi}\int_0^{\infty} q^2dq\ \frac{j_0(qr)}
{q^2+\mu^2}\ v^2(q). 
\qe
The expressions Eqs.~(\ref{reidpot},\ref{ope1}) above result from 
taking
\eq
v^2(q)=\frac{M^2-\mu^2}{q^2+M^2}.\label{v2}
\qe

To compare with values of $M$ taken from the literature 
some correspondence must be made. In no case is an exact 
comparison possible. Often one compares form factors by 
making an expansion for small $q$ and matching the 
coefficients of $q^2$. While it is not clear that this is 
the proper procedure, since it corresponds to comparing 
at large $r$ and we are interested in the behavior at 
small $r$, we will often follow that method as well.

The pion form factor reduces the strength of the potential 
at small values of $r$. For the form mentioned above

\eq 
\frac{1}{\mu^2+q^2} \frac{M^2-\mu^2}{M^2+q^2} 
\rightarrow \frac{\pi}{2}\frac{(e^{-\mu r}-e^{-M r})}{r}.
\qe

The proper value of the pion-nucleon form factor has been much 
debated over the years 
(\cite{layson,landtab,doverff,lmm,ernstff,thomasff,kukulin}.
A survey of this early literature leads us to values of
$M$ is the range of 1.4 to 2.8 \fmm.

Recently there has been a great deal of interest in the 
development of an effective field theory in which the 
dependence on the form factor range would disappear at the cost 
of a renormalization order by order. While this program has met 
with partial success \cite{marji}, an observable dependence on 
the range remains.

In a recent high-quality fit to the NN phase shifts, Entem and 
Machleidt \cite{entem} used a Gaussian form factor with 
$\Lambda_g=0.5$\ GeV. When using potentials of this type to 
construct the Equation of State of neutron matter 
\cite{sam1,sam}, Gaussian form factors with a range for 
$\Lambda_g$ of 450 MeV/c $\le \Lambda_g \le$ \ 650 MeV/c were 
used. Epelbaum, Glockle and Meissner used $\Lambda_g$ around 
600 Mev/c in N$^3$LO calculations \cite{epel2}. Epelbaum and 
Meissner suggest \cite{epel1} that $\Lambda_g\approx 3$ \fmm\ 
is the value that should be used. Values of $\Lambda_g$ in the 
same range appear optimal in the strange sector 
\cite{polinder,haiden,gazda} as well.

The form factor is often regarded as the Fourier transform of 
the density of elementary scattering centers in a composite 
target. With the assumption of a Gaussian density the form 
factor becomes 

\eq S(q)=e^{-<r^2>q^2/6}=e^{-q^2/\Lambda_g^2}. 
\qe 

Using $<r^2>^{\h}=0.86$ fm (a common value for the charge 
radius of the proton) leads to $\Lambda_g=562$\ MeV/c.

To attempt to match these Gaussian form factors to the one we 
have used, one can consider several possibilities. From the 
expansions for small $q$ one gets $M=\Lambda_g/\sqrt{2}$. One 
might also choose to make the two form factors (or their 
squares) equal at some momentum scale $q_0$, i.e., 
\eq 
\frac{1}{1+q_0^2/M^2}=e^{-2q_0^2/\Lambda_g^2} \qe
Choosing $q_0=M$ we find $M/\Lambda_g =\sqrt{\ln{2}/2}\approx 
0.588$ leading to $M=1.79$ \fmm for $\Lambda_g=600$ MeV/c. 

We remind the reader that there exists in the literature a 
least one example \cite{rho} where it was pointed out 
\cite{wolo} that leaving out the pion-nucleon form factor in 
the calculation of the meson-exchange-current contribution to 
elastic electron scattering led to a wrong prediction (by a 
factor of 30) as verified by experiment.

Thomas and Holinde \cite{thomasholinde} pointed out that the 
apparent discrepancy between the charged and neutral $\pi NN$ 
coupling constants is resolved if the mass in a monopole for 
each vertex is in the range 500-800 MeV (corresponding to 
$M\approx 350-560$ MeV or $\approx 1.75-2.80$ \fmm.

In summary, the values quoted here from the literature lie mainly 
in the range $M=260-500$ MeV/c (1.4 to 2.8 \fmm). We adopt a value 
of $M=2$ \fmm\ as representative of these limits.

\section{Extracting the ``Coulomb corrected'' $pp$
phase shifts\label{extract}}

We now give the method used to extract the pure $pp$ singlet 
$S$-wave phase shifts from the data. By adding the closed form for 
the Rutherford amplitude 
\eq 
f_R(\theta )=-\frac{\eta}{2k\sin^2\frac{\theta}{2}}
e^{i(2\sigma_0-\eta\ln \sin^2\frac{\theta}{2})} \label{ruth}, 
\qe

and subtracting the corresponding 
partial-wave expansion, we can write the full amplitude as
\begin{eqnarray}
f(\theta )&=&f_R(\theta)+\frac{1}{2ik}\sum_{\ell}(2\ell+1)P_{\ell}
(\cos\theta )\left( e^{2i\delta_{\ell}}-e^{2i\sigma_{\ell}}\right) \label{ampexc} \\
&=&f_R(\theta)+\frac{1}{2ik}\sum_{\ell}(2\ell+1)P_{\ell}
(\cos\theta )e^{2i\sigma_{\ell}}
\left( e^{2i\delta_{\ell}^e}-1\right),
\label{ampexc1}
\end{eqnarray}
where $\delta_{\ell}$ in Eq.~(\ref{ampexc}) is the full phase shift obtained by 
matching the solution of the Schr\"odinger equation with the 
strong plus CC and standard Coulomb potential 
included. The $\sigma_{\ell}$ are the ``Coulomb phase shifts'' 
which result from the solution of the Schr\"odinger equation with 
a point Coulomb potential. Since for large $\ell$, $\delta_{\ell} 
\rightarrow \sigma_{\ell}$, this series converges. 

In equation~(\ref{ampexc1}) the modified phase shifts $\delta_{\ell}^e$ (called the ``electric phase shifts'' by 
Heller \cite{heller})
\eq \delta_{\ell}^e\equiv\delta_{\ell}-\sigma_{\ell}. \qe
go to zero for large $\ell$ and at
zero energy where both $\delta_{\ell}$ and $\sigma_{\ell}$ diverge. \\ 

\begin{figure}[htb]
\epsfig{file=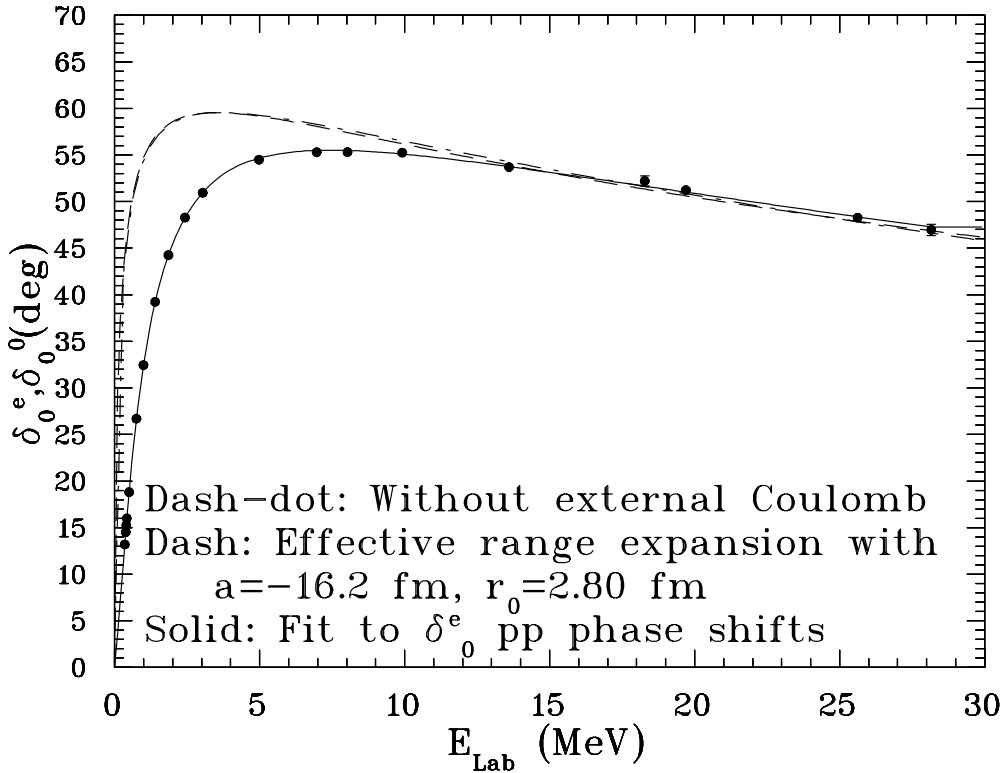,angle=90,height=4.in}
\caption{Typical phase shift fit with the Reid-like potential 
(curve), [including the CC potential] to the values 
from Ref \cite{ppdat} (solid dots).  This fit corresponds to the 
solid line in Fig. \ref{potpps}. The difference between the dash-dot
and solid curves reflects the importance of the electric correction.
The fitted model has $p=3$ fm$^3$ and $\beta=4$~\fmm.
\label{phasereid}} 
\end{figure}

We are interested in the phase shifts, $\delta_{\ell}^0$, given 
by the strong interaction alone and $\delta_{\ell}^{cc}$, which 
results from the strong plus CC potential. It is the latter which 
should be used in the calculation of nuclear properties and the 
former which are to be considered as ``isospin symmetric'' 
results. In the limit that the standard Coulomb interaction goes 
to zero (so that $\delta_{\ell}^e\rightarrow 
\delta_{\ell}^{cc}$), this simple addition of phase shifts might 
be considered as the first order correction to the phase shift 
obtained with only the strong + CC interaction for the standard 
Coulomb effect, although $\delta_{\ell}^e$ clearly contains a 
further dependence on the standard Coulomb interactions. In order 
to calculate this additional dependence, models for the strong 
interaction are used. This correction is sometimes called the 
``inner Coulomb correction'' in pion-nucleon scattering but we 
will call it the ``electric correction'' to avoid possible 
confusion with the CC effect. It only takes into account the 
Coulomb explicitly included in the Schr\"odinger equation, of 
course. Any hidden Coulomb effect (such as the crypto-Coulomb 
that we consider here) would not figure in this correction; it 
would appear as being part of the strong interaction.

The quantity $\delta_{\ell}^e$ can be obtained directly from the 
$pp$ data without recourse to any model \cite{ppdat}. For 
this reason it makes a good contact point for fitting models to the 
data. The precision of the data is among the best in nuclear physics: 
The low-energy points have uncertainties below 1 \%, most less that 
0.1\% while those above 5 MeV have uncertainties of the order of 0.2 
to 1\%. We have added (in quadrature) uncertainties of 0.1\% to the 
quoted values to take into consideration the crudeness of our model. 
We find that the data points are fit to within the order of 0.1\% 
without difficulty (see Fig \ref{phasereid}).

Turning off the entire Coulomb potential, including the CC part, we 
obtain $\delta_0^0$ and from it, the $a_0$ (isospin unbroken) 
scattering length. Keeping only the strong plus $pp$ CC potential we 
obtain the $pp$ version of $\delta_0$ and from it the value of 
$a_{pp}$ is obtained. By adding the attractive CC $np$ potential (see 
Fig.~\ref{potpns}) to $V_0$, $a_{np}$ is obtained and with the 
repulsive $nn$ potential (Fig \ref{potnns}) $a_{nn}$ is found. This 
procedure was followed for the case of no pion mass difference and 
the physical difference between charged and neutral pion masses. The 
entire process was carried out for 3 different sizes of the diquark 
correlation range as expressed by $\beta$ and a sampling of values of 
the parameter $p$.

Figure \ref{phasereid} shows a fit (solid line) to the data. 
The dash-dot line shows the results
of a calculation using the strong + CC potential only. One sees
that above 15 MeV the electric correction is very small, one
would have obtained nearly the same values of $\delta_0^{cc}$ 
from a direct comparison with the data for $\delta_0^e$.
The result of the effective range formula
\eq
k\cot \delta_0=-\frac{1}{a}+\h r_0 k^2.
\qe
is also given (dashed line) and it is seen that the representation is very
good. For $a=-16.2 $ fm and $r_0=2.80$ fm, the first term contributes 
0.062 \fmm  and (at 15~MeV) the second gives 0.25 \fmm\ so that 
the effective range term dominates ($\ge 80\%$) at these energies 
and the fit is much more sensitive to $r_0$ than to $a$. Of 
course, at the very low energies the reverse is true. 

\section{Results\label{results}}
\begin{figure}[htb]
\epsfig{file=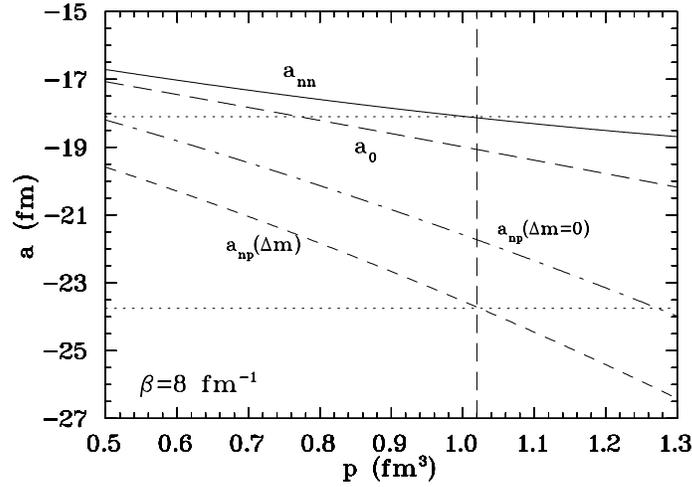,angle=90,height=2.5in}
\caption{Scattering length vs. the strength parameter $p$ for 
$\beta=8$\ \fmm. The notation
$\Delta m=0$ means that the difference in pion masses is not
taken into account, the neutral pion mass is used in all one-pion
exchange calculations. The notation $\Delta m$ means that the
charged and neutral pion masses are used as in Eq. \ref{ope1}.}
\label{slvsp8}
\end{figure}

\begin{figure}[htb]
\epsfig{file=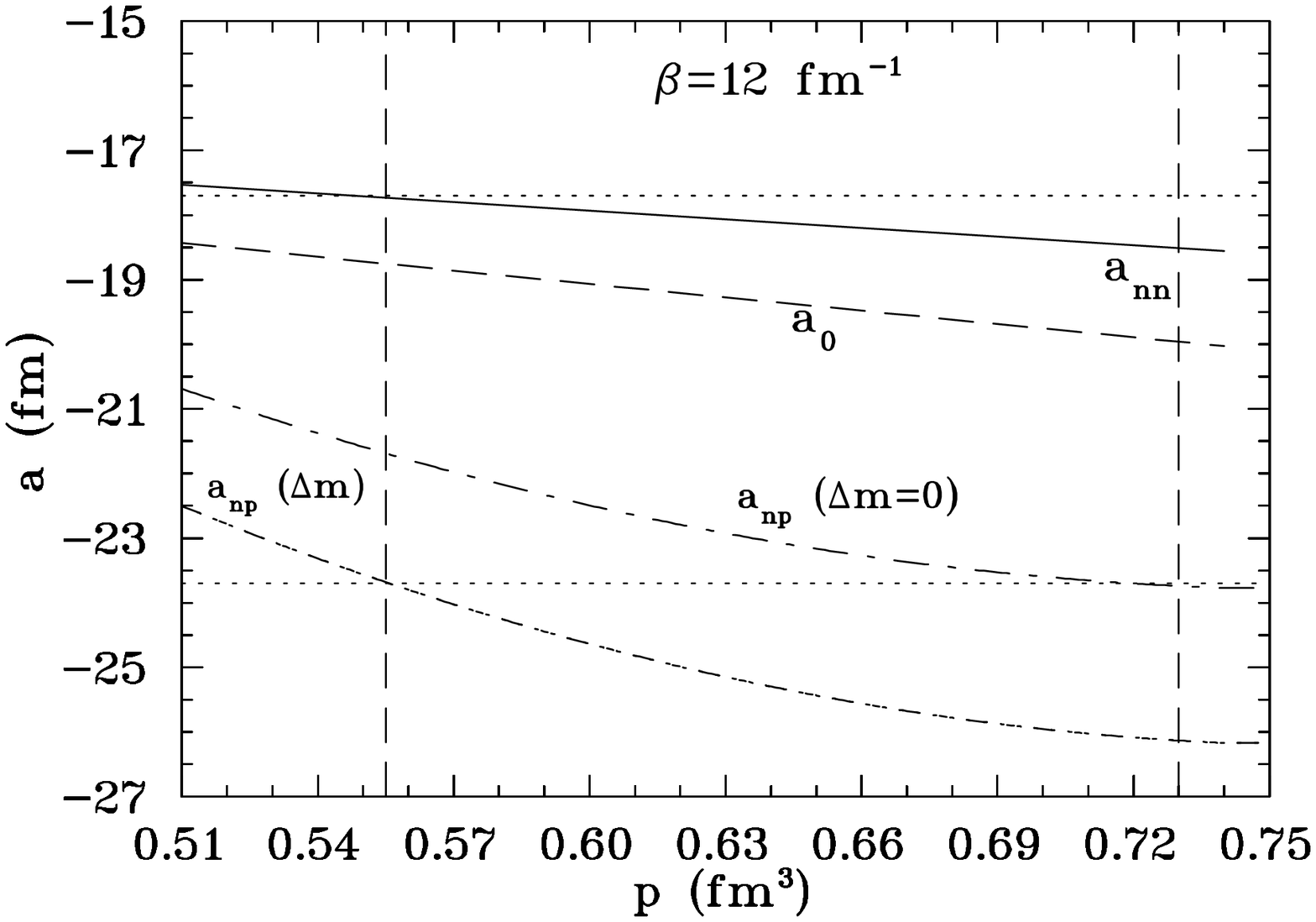,angle=90,height=2.5in}
\caption{Scattering length vs. the strength parameter $p$ for 
$\beta=12$\ fm$^{-1}$. See Fig \ref{slvsp8} for notation.}
\label{slvsp12}
\end{figure}

\begin{figure}[htb]
\epsfig{file=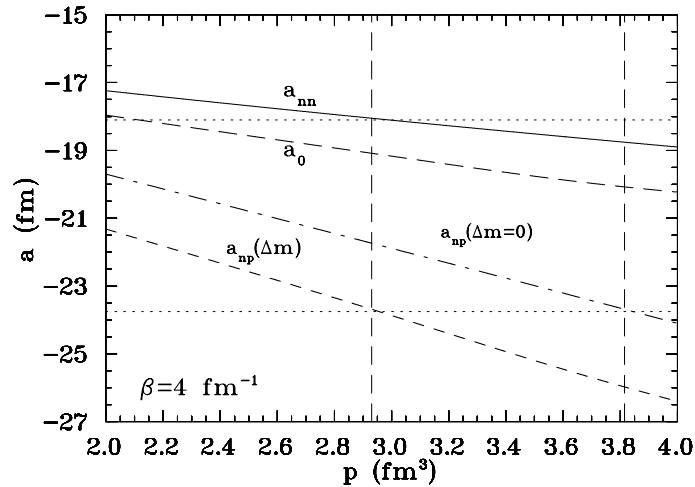,angle=90,height=2.5in}
\caption{Scattering length vs. the strength parameter $p$ for 
$\beta=4$\ fm$^{-1}$. See Fig \ref{slvsp8} for notation.}
\label{slvsp4}
\end{figure}
q
\subsection{Relationship among the scattering amplitudes}

The curves in Figs. \ref{slvsp8}, \ref{slvsp12} and \ref{slvsp4} show 
how the amplitudes are related to each other and were obtained by the 
following procedure: The experimental $pp$ phase shifts, 
$\delta_{\ell}^e$ taken from reference \cite{ppdat} were fit using an 
isospin invariant strong potential (including one-pion exchange, 
$V_0(r)$ of Eq. (10) plus the full Coulomb potential (some examples are 
shown in Fig. \ref{potpps}) generated by the model for a set of values 
of the parameter $p$. As $p$ is varied the coefficients $C_1$ and $C_2$ 
must also change in such a way as to keep $V_0(r)$ plus the CC 
potential the same so as to fit the data. When the ordinary Coulomb 
potential, $V_C(r)$, is removed, as in the standard procedure, the 
resulting $a_{pp}$ calculated from this combined potential will be the 
same, or very nearly so. This potential would be considered as the 
``strong potential'' if one were not aware of the presence of the CC 
potential. This compensation of variation of $V_0(r)$ and the CC 
potential it seen to be very good since the values of $a_{pp}$ are 
found to be (very nearly) independent of $p$ at $-16.2$~fm.

As $p$ varies, however, $C_1$ and $C_2$ {\it do change} so that when 
$V_0(r)$ alone is used to calculate the pure isospin conserving 
scattering length, $a_0$, does change, as do $a_{nn}$ and $a_{np}$ 
when calculated with their respective CC potentials.

To find the $nn$ scattering length $a_{nn}$ in these figures follow 
the short-dash curve until it crosses the experimental value of 
$a_{np}$ (lower horizontal dotted line) to obtain the value of $p$ 
appropriate. Then the value of $a_{nn}$ can be read off at that value 
of $p$, for example in figure \ref{slvsp8}, $p=1.02$\ fm$^3$, 
$a_{nn}=-18.1$\ fm.

\begin{figure}[htb]
\epsfig{file=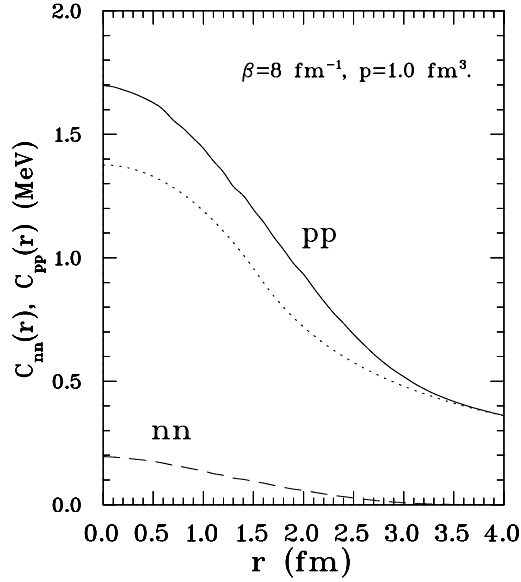,height=3.in}
\caption{Neutron-neutron and $pp$ potentials 
resulting from the calculations described in the text. The 
$pp$ case includes the normal Coulomb interaction as well as 
the hidden part. The dotted line shows the potential from a 
uniformly charged sphere of radius 1.56 fm, $V_C(r)$.
In order to find the CC part the differences in these curves 
is taken. \label{csb8a}}
\end{figure}

\begin{figure}[htb]
\epsfig{file=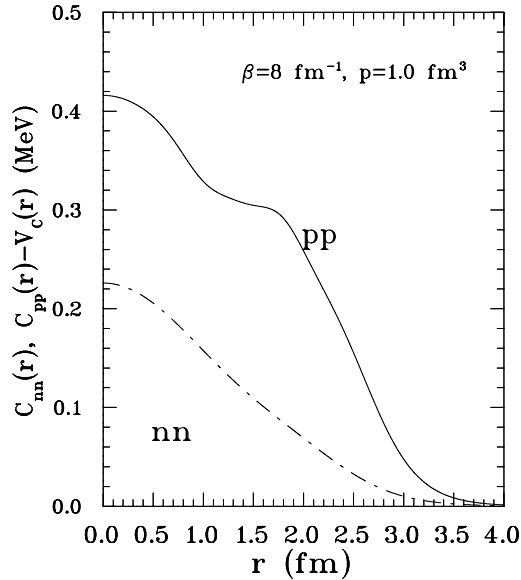,height=3.in}
\caption{Crypto-Coulomb potentials with
the potential from a uniformly charged sphere of radius 1.56 
fm subtracted from the $pp$ interior potential.
\label{csb8b}}
\end{figure}

\begin{figure}[htb]
\epsfig{file=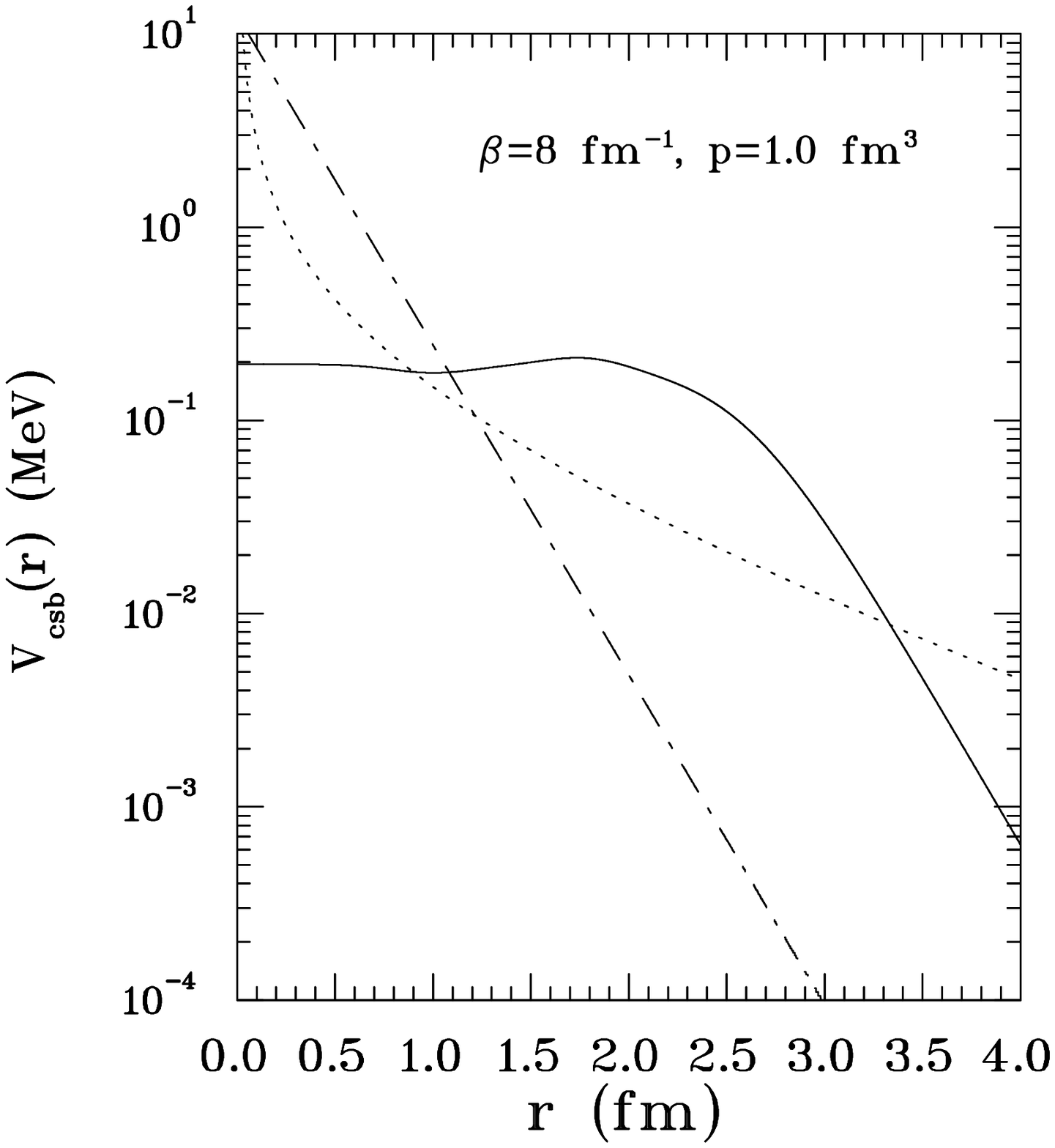,height=3.in}
\caption{CC charge symmetry breaking potential
calculated by subtracting the uniformly charged  potential and 
the $nn$ interior Coulomb potential from the $pp$ interior potential
(solid line).
The dotted line shows the shape of a one-pion-exchange potential
with arbitrary normalization. The dash-dot line shows the first
term of the $\rho$-$\omega$ CSB potential (Equation 8 of ref. 
\cite{sid} with their parameter $\beta=0$).\label{csb8cl}}
\end{figure}

\subsection{Charge-symmetry-breaking Potentials}

The range of the correlation $r_{corr}=\sqrt{12}/\beta$ is 0.867\ 
fm for $\beta=4$\ 
\fmm, 0.433 fm for $\beta=8$\ \fmm \ and 0.289 fm for $\beta=12$ \fmm \  
compared with 0.27 fm from Bloch et al. \cite{bloch}, 0.32 
fm from Carlson, Kogut and Pandharipande \cite{carlson1}, 0.21 fm
from Sisodiya et al. \cite{sisodiya},  
and 0.24 fm from Keiner \cite{keiner}.

While it may at first appear that the order of the scattering
lengths has changed since the $nn$ scattering length is now less 
negative than the unbroken one, the effect is the same as before
since the CC potential for the $pp$ system is more 
repulsive than that of the $nn$ system so that there is an 
apparent stronger Coulomb effect in (say) the three-body system.
One can take directly the difference of these potentials to find
a charge symmetry breaking potential which could be used in the
same way as the meson mixing potentials~\cite{sid}.

In general these potentials could be used directly in any 
calculation of nuclear properties. However, a charge symmetry 
breaking potential to be used in the same spirit as the 
$\rho-\omega$ potential discussed earlier in a perturbation 
calculation can also be obtained by removing the ``ordinary'' 
Coulomb potential (since the desired potential is to be beyond 
the standard Coulomb calculation) and then removing the 
interior repulsive potential of the $nn$ case. Figures 
\ref{csb8a}, \ref{csb8b} and \ref{csb8cl} show this progression 
and selected analytical fits to the breaking potentials are 
given in the appendix.

\subsection{$^3$He-$^3$H mass difference}

Several cases have been calculated with our variational wave 
function introduced in Ref. \cite{gd}. 

\begin{table}[htb]
$$\begin{array}{ccccc}
&\beta={\rm 4\ fm^{-1}}&\beta={\rm 8\ fm^{-1}}&\beta={\rm 12\ 
fm^{-1}}&{\rm Experimental\ value}\\
{\rm E_{^3H}\ (MeV)(With\ 
crypto{\rm -}Coulomb)}&8.5034&8.4966&8.4999&8.4818\\
{\rm E_{^3H}\ (MeV)(Pure\ Isospin)}&8.5475&8.5452&8.5490&\\
{\rm Difference\ (keV)}&44.1&48.6&49.1\\
{\rm E_{^3He}\ (MeV)(With\ 
crypto{\rm-}Coulomb)}&7.7869&7.7760&7.7770&7.7181\\
{\rm E_{^3He}\ (MeV)(Charged\ sphere)}&7.8810&7.8787&7.8825&\\
{\rm Difference\ (keV)}&99.4&102.7&105.5\\
{\rm Difference\ between\ ^3He\ and\ ^3H\ 
(keV)}&716.5&720.1&720.6\\
{\rm Including\ 46\ keV}&762.5&766.1&766.6&763.8\\
\end{array}$$
\caption{Coulomb differences in the 3-nucleon system.
The 46 keV in the last line comes from the estimate of several
small effects by Wu, Ishikawa and Sasakawa \cite{wu} and
Friar, Gibson and Payne \cite{friar1}.}
\label{threebody} \end{table}

In order to calculate with a realistic wave function we use a trial
wave function which results from a variational calculation for 
$^3$H. Since the true wave function must be translationally
invariant it can depend only on the relative coordinates,
$\bfr_{ij}=\bfr_i-\bfr_j$.

We choose the basic form
\eq
\psi(\bfr_1,\bfr_2,\bfr_3)=f(r_{12})f(r_{13})f(r_{23})
\qe
where $r_{ij}=|\bfr_{ij}|$ and the function, $f(r)$ is arbitrary at
this point. It will be chosen with a number of parameters to be
selected to minimize the energy. 

We originally took for the form of $f$
\eq
f(r)=(1-e^{-c'r})\frac{e^{-ar}}{b+r}
\qe

the first factor vanishes at the origin and provides the effect 
of a (very mild) repulsive correlation. The rest of the function 
has the proper asymptotic form. For the relatively long-range 
isospin breaking potentials that we have used before, the very 
short-range region was not crucial but with the full NN 
potential, with its strong repulsion, we need to model the 
short-range exclusion region of the factors. To this end we 
include an additional factor in $f(r)$ which is \eq 
e^{-\alpha_n/r^n}, \qe where both $n$ and $\alpha_n$ can be 
treated as variational parameters.

The $np$ interaction was modified so as to obtain approximately
the correct binding energy for $^3$H by decreasing the repulsive
core by 38\%. Since we have set the parameters of the wave function
to give the correct three-nucleon r.m.s. radius (we deem the range
over which the potential is tested to be important) this choice
of an $np$ potential to give the correct binding energy is only
cosmetic. The mass difference in the three-nucleon system could
be obtained directly by the expectation value of the breaking
CC potentials.

See table \ref{threebody} for a summary of the results. The 
difference in energy between the conventional Coulomb and the 
crypto-Coulomb for $^3$He is about 100~keV. Of course the Triton 
also has an effect from the $nn$ repulsion which, overall gives 
about 50~keV less binding with the CC so the 
difference beyond the conventional Coulomb is near 50 keV which 
is about what Wu, Ishikawa and Sasakawa \cite{wu} and Friar, 
Gibson and Payne \cite{friar1} find is needed after a number of 
small effects have been taken into account.

\section{Extension to the strange sector\label{strange}}

Similar considerations can be applied to the $\Lambda n$ and
$\Lambda p$ systems. In the diquark picture the $\Lambda$ consists
of quarks $ud$ coupled to I=0 with charge +1/3 and the strange
quark with charge -1/3. Thus it is electrically the same as the
neutron. If we were to assume that the masses and the interactions
among the quarks were all the same, SU(3), then we could directly
use the $nn$ and $np$ crypto-Coulomb potentials obtained previously. 

\begin{table}[htb]
$$\begin{array}{lccccr}
\hline
{\rm case}&\Lambda p\ a&\Lambda n\ a &\Lambda p\ r_0&\Lambda 
n\ r_0 &\%\ diff\ a\\
&{\rm fm}&{\rm fm}&{\rm fm}&{\rm fm}\\
\hline
OPE=0 &  -4.17  &  -3.89 &  2.88 &   2.91   & 6.7\\
C_1\times 0.9535 &  -2.51  &  -2.36 & 3.46 &   3.51   & 6.0 \\
C_2\times 1.0785 &  -2.51  &  -2.37 & 3.53 &   3.58   & 5.5 \\
V\times 0.882    & -2.51   &  -2.37 & 3.33 & 3.37     & 5.5 \\
\hline
{\rm NSC97f}\ &  -2.51&  -2.68&  3.03&  3.07&  -6.8\\
{\rm ESC08c}\ &  -2.46&  -2.62&  3.14&  3.17&  -6.5\\
{\rm NSC89}\ &  -2.73  &-2.86&  2.87&  2.91&  -4.7\\
\hline
\end{array}$$\label{lmb}
\caption{SU(3) predictions for singlet $\Lambda N$ scattering parameters.
The calculations were made with NN potentials which give $a_{np}=-23.54$ 
fm, $a_{nn}=-18.09$ fm, $r_0(np)=2.69$ fm and $r_0(nn)=2.82$ fm. The last
two lines are from fits by the Nijmegen group.  NSC97f is from Rijken et al. 
\cite{rijken}, ESC08c is from Nagels et al. \cite{nagels}, and NSC89 is
from Maessen et al. \cite{nsc89}. 
Our calculations are for $p=1.0$ fm$^3$ and $\beta=8.0$
fm$^{-1}$.}
\end{table}

Table IV shows the results of several variations. The first line 
shows the result of removing the one-pion-exchange potential. We see 
already that the right order of magnitude is obtained. Then, by reducing 
the strength of the remaining potential in three different ways one can 
match the NSC97f $\Lambda p$ result of Rijken et al. \cite{rijken}.

Rijken et al. \cite{rijken} give six fits to hyperon-nucleon scattering 
data of equivalent quality labeled a-f. The solution $f$ is emphasized in 
the paper and their G-matrix calculation of hyperonic systems also favors 
it. Miyagawa et al. \cite{miya} give limits on the singlet and triplet 
scattering and only the $f$ solution falls within these limits. In 
Ref.~\cite{gchg} the singlet and triplet scattering lengths were 
extracted from the results of a feasibility experiment of radiative Kaon 
capture on deuterium. While the limits on the singlet scattering length 
are too large ($-0.15\rightarrow -5.0$ fm) for a comparison to be useful, 
the triplet determination ($-1.3\rightarrow -2.65$ fm) does provide some 
constraint and is in agreement with the triplet version (not shown) of 
all of the potentials corresponding to the scattering parameters shown in 
the last three lines of Table IV. The principal isospin breaking 
mechanism used in the potentials of Rijken et al. \cite{rijken} (and 
others) is $\Lambda \Sigma$ mixing, first introduced by Dalitz and von 
Hippel (DvH) \cite{dalitz}.

The potential NSC89 \cite{nsc89} deserves further discussion. It has a 
larger $\Lambda \Sigma$ mixing in the few-baryon system and so lends 
itself to the understanding of the anomalous $\pi^+$ weak decay of 
$_{\Lambda}^4$He \cite{benrob}. However, perhaps partly due to the fact 
that it gives twice the excitation energy of the $1^+$ state in 
$_{\Lambda}^4$He than that observed, it seems to have been rejected by the 
community and it has been said that ``NSC89 is definitely not a realistic 
YN potential for use in hypernuclei \cite{galrmp}.''

The potential ESC08c comes \cite{esca,escb,nagels} from a set of 
potentials which use the forbidden state concept to implement the Pauli 
blocking at short distances and includes explicit two-pion exchange.

We see that Nijmegen breaking potentials are more attractive for the 
$\Lambda n$ interaction than for the $\Lambda p$ interaction in the 
singlet channel. The opposite is true for the triplet channel. This 
behavior is expected for the $\Lambda \Sigma$ mixing mechanism for the 
breaking as by the formula given by Gal \cite{gal}. Gal gives 
\cite{gal,gazda} a simple relation of the breaking potential to the strong 
interaction potential causing the transition $\Lambda\rightarrow\Sigma$. 
The interaction has a major component of one-pion exchange (OPE).

 The OPE potential consists of a central (spin-spin) and a tensor term. If 
only the spin-spin term were present, the singlet to triplet ratio of the 
breaking potentials would be -3:1. The tensor part contributes (only in 
the triplet state) and alters this ratio. For the Nijmegen triplet 
scattering lengths (not shown) we estimate that the ratio of the 
corresponding breaking potentials is about -1:1.

The percentage breaking observed for the triplet state is about the 
same as the singlet but in the same direction as the CC effect. 
Assuming that both DvH and CC potentials contribute, we would expect 
that they tend to cancel in the singlet and add constructively in the 
triplet. This would lead to a large breaking ($\approx 12 \% $) in 
the triplet state, a fortunate circumstance since the $\Lambda n$ 
triplet appears \cite{gchg} to be easier to extract from data than 
the singlet.

As in the non-strange sector, there are tests available in
mirror hypernuclei, the most common being the ground ($0^+$) and
first excited states ($1^+$) of $_{\Lambda}^4$H and $_{\Lambda}^4$He.

The experimental data come from several sources which are 
enumerated in Yamamoto et al. \cite{yama}. The ground-state 
binding energies were measured with an emulsion technique by 
Juri\'c et al. \cite{juric} $B^{0+}(_{\Lambda}^4He)=2.39\pm 0.03$ 
MeV and $B^{0+}(_{\Lambda}^4H)= 2.04\pm 0.04$ MeV. A modern 
measurement was presented by Esser et al. \cite{esser}, 
$B^{0+}(_{\Lambda}^4H)=2.12\pm 0.09$ MeV. The binding energy of 
the $1^+$ excited state is obtained by adding the energy 
difference obtained from the measured energy of the $\gamma$ from 
the transition. Yamamoto et al. \cite{yama} (aside from their own 
measurement of the transition energy in 
$_{\Lambda}^4$He$=1.406\pm 0.003$) obtained the $\gamma$ energy 
for $_{\Lambda}^4$H$=1.09 \pm 0.02$ MeV from the average of three 
experiments \cite{bed1,bed2,kaw}. The breaking data are summarized 
in table V. Experimental column $a$ is calculated using the 
ground-state measurements of Juri\'c et al. \cite{juric} while 
column $b$ substitutes the measurement of Esser et al. 
\cite{esser} for $B^{0+}(_{\Lambda}^4$H). Column $c$ shows the 
breaking in the excitation energy from Yamamoto et al. 
\cite{yama}.

Nogga et al. \cite{nogga} addressed the challenges of choosing the 
``best'' hyperon-nucleon interaction by investigating the set of 
potentials mentioned above \cite{rijken}. They found that NSC97f 
under-binds $_{\Lambda}^3$H (but by considerably less than the rest of 
the 97 set) while NSC89 slightly over-binds the three-baryon system.
Mirror-nuclei tests are not available in this system.

For the difference in energy between the ground state and the $1^+$
excited state in the four-body system they found that NSC97f gives the
correct value although the absolute binding is too small. No potential was
found to give the correct values for all observables.

They also studied the the contribution of the spin to the total angular
momentum and found that the ground state is about 90 \% S=0 and the
first excited state is 96\% S=1. If one were to take the like-nucleon
pair to be coupled to spin zero in a pure s-wave model, the $\Lambda$-odd
nucleon pair would determine the spin of the total system. In this case
the ground state would be pure $\Lambda-N$ singlet and the excited state
pure triplet. The percentages found by Nogga et al. \cite{nogga} are
reasonably consistent with this simple picture.

Nogga et al. \cite{nogga} also calculated the effect of the CSB on the 
difference in separation energies of the $_{\Lambda}^4$H and 
$_{\Lambda}^4$He systems. They found that NSC97e gives only 70 keV of the 
experimental value of 270-350 keV. On the other hand NSC89 gives 
essentially all that is needed.

We see that our results show a more attractive $\Lambda p$ than $\Lambda 
n$ potential reflected in the fact that the scattering length is more 
negative. This is in the direction needed since the separation energy of 
the $\Lambda$-$^3$He system (2.39) Juri\'c et al. \cite{juric} (see also 
Esser et al. \cite{esser} is greater than that of the $\Lambda$-$^3$H 
system from Juri\'c et al. \cite{juric} (2.04 MeV) or Esser et al. 
\cite{esser} (2.12 MeV). The singlet scattering lengths of Rijken et al. 
\cite{rijken} show the opposite behavior. The crypto-Coulomb attraction 
in the $\Lambda p$ system and the repulsion in the $\Lambda n$ system 
can be expected to be about the same in the triplet as the singlet since 
their origin is the correlation of the odd quarks of different or same 
charge.

\begin{figure}[htb]
\epsfig{file=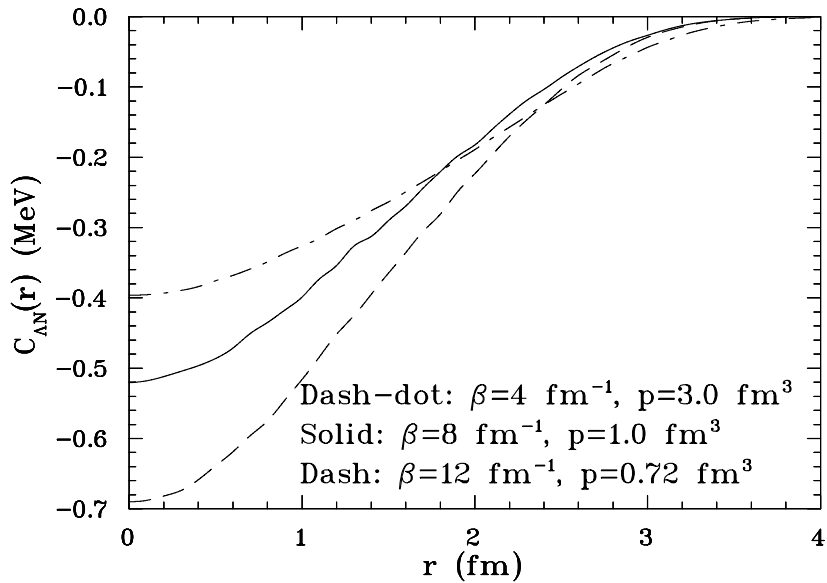,angle=90,height=3.in}
\caption{Three examples of $\Lambda N$ breaking potentials from
Eq. \ref{breakpot}\label{potslam}}
\end{figure}

Since the resulting charge symmetry breaking potential 
\eq C_{\Lambda N}(r)\approx C_{np}(r)-C_{nn}(r)\label{breakpot}
\qe
is of the order of 300-400 keV (see Fig. \ref{potslam} and the difference 
of the separation energies is 270-350 keV, it seems that one is bound to 
find a significant fraction of this difference from this source. From 
general considerations in the present model we see that CSB in the strange 
sector (by the measure of the separation energies) corresponds to CIB in 
the non-strange sector.

\begin{table}[htb]
$$\begin{array}{lcccc|ccc}
\hline
\Lambda_g({\rm MeV})\ &550 &600&650&700  &{\rm Exp\ (a)}&{\rm 
Exp\ (b)}&{\rm Exp\ (c)}\\
\hline
\Delta^{1+}=
B^{1+}(_{\Lambda}^4He)-B^{1+}(_{\Lambda}^4H)
&-172&-193&-228&-223&+30\pm 50&-50\pm 90\\
\Delta^{0+}=B^{0+}(_{\Lambda}^4He)-B^{0+}(_{\Lambda}^4H)
&30&136&244 &294&350\pm 50&270\pm 130\\
\Delta^{0+}-\Delta^{1+}&202&329&492&517&&&316\pm 20\\
\hline
\end{array}$$
\label{tbl4}

\caption{Isospin breaking energy differences as a function of cut-off 
parameter taken from Gazda and Gal \cite{gazda} (columns 2-5) compared 
with measurements (columns a-c). The experimental energy breaking 
difference between the ground and first excited state (column c) is taken 
directly from the measurements of the difference in the $\gamma$ energies 
\cite{yama}. See the text for references to the data.}\end{table}

An important contribution to the understanding of CSB in the 
four-body system was recently reported by Gazda and Gal 
\cite{gazda}. They present the results of no-core shell model 
calculations of $_{\Lambda}^4$He and $_{\Lambda}^4$H for the 
ground state ($0^+$) and first excited state ($1^+$). They use a 
CSB interaction from the DvH \cite{dalitz} mechanism which has 
(largely) pion coupling to mix the isospin-zero $\Lambda$ with 
the isospin-one $\Sigma^0$. They use YN potentials from Polinder 
et al. \cite{polinder} calculated with effective field theory to 
leading order and available for a range of values of cut-off 
mass. They find a strong dependence on the cut-off mass (see 
Table V). Restricting ourselves to the lower values of 
$\Lambda_g$\ we see that the breaking in the ground state is 
compatible with the 70 keV found by Nogga et al. \cite{nogga}, 
much smaller than the experimental value of 270-350 keV.

The breaking in the difference of the energies is equivalent to the 
breaking in the directly observed $\gamma$ rays \cite{yama}. The breaking 
in the excited state is calculated \cite{gazda} to be around --200 keV 
whereas the experimental value is around zero. A constant shift of the 
breaking in both states ($0^+$ and $1^+$) of $\approx$ 200 keV would give 
agreement with data. An energy shift of 200 keV is about the result 
expected from the CC mechanism presented here since, from arguments given 
earlier, we expect little or no spin dependence for the CC mechanism.

Thus the mixing (DvH) mechanism gives attraction for the ground state
and repulsion for the excited state. This is perhaps understandable
since the spin-spin coupling which is providing the mixing changes
sign between singlet and triplet states in the usual manner.

It may be useful to discuss the possibly confusing result that the 
maximum breaking in the scattering lengths occurs in the triplet 
($1^+$) state while the maximum breaking in the bound state is in the 
singlet ($0^+$) state. The CC effect is attractive for the $\Lambda p$ 
interaction and independent of spin so it increases the binding in 
$_{\Lambda}^4$He (relative to $_{\Lambda}^4$H) in both states. It was 
pointed out by Gal \cite{gal} that the two like nucleons in the 
three-body core are paired to spin zero, hence do not interact with the 
mixing pion exchange leaving only the odd nucleon to cause the mixing. 
Since, with the DvH mechanism, the $\Lambda n$ breaking interaction is 
more attractive in the singlet state, its interaction with the single 
neutron in the nucleon core leads to an attractive interaction in the 
$_{\Lambda}^4$He case and it adds to the CC potential constructively. 
Since the DvH potential changes sign for the triplet case, the two 
potentials have opposite signs, leading to a smaller breaking
 
It may be useful to use the breaking in the energy of the $\gamma$
between the two states as a constraint \cite{gazda} since it has the
smallest experimental uncertainty. Doing that, in Table V we would
choose $\Lambda_g=600$ MeV to compare. Adding 200 keV to the two
bindings leads to rough agreement with the data.

There is one caveat in the comparison with the results of Gazda and Gal 
\cite{gazda}. Their calculation was done with the fit to the leading 
order effective field theory \cite{polinder}. Doing the same calculation 
with the next to leading order as input \cite{haiden} leads to results 
inconsistent with data \cite{gazda}.

\section{Discussion\label{discuss}}

The general idea presented is that quark clustering will lead to isospin 
breaking through the Coulomb interaction. We have attempted to present 
this notion with a very simple illustrative model based solely on valence 
quarks. Since the breaking interaction arises from the Coulomb potential, 
it addresses CSB and CIB on an equal footing. Of course, the breaking due 
to the pion mass difference contributes to the CIB as well although we 
find that its influence is considerably reduced due to form factors.

If it seems to the reader that we are attacking the problem of the 
low energy $NN$ interaction by starting in the middle, it is because 
we are. Our object is to show that it is possible to establish a link 
between the quark rearrangement, which must occur in any complete 
theory of $NN$ scattering and isospin breaking. We have attempted to 
do so by considering a simple model, in which we included what we 
believe to be the basic physics of the system. Since only the quark 
density is needed to calculate the Coulomb energy many of the details 
of the wave function are unnecessary, allowing progress to be made 
with a minimum of assumptions.

The meson-nucleon form factor range plays a significant role in the 
understanding of CSB and CIB. The value of the regulating mass that we 
have adopted may appear small to some but we believe we have justified 
it with a number of references. It is interesting to note that, in the 
effective field theory approach, both in the non-strange 
\cite{entem,sam1,sam} and strange sectors \cite{polinder,haiden} the 
minimum $\chi^2$ is obtained for values of $\Lambda_g$ around 500-600 
MeV/c, in general agreement with previous determinations of the cut-off 
cited in Section IV D and also compatible with the value derived from 
the size of the nucleon in the same section.

The value that we obtained for the magnitude of the ``Coulomb 
corrected'' $pp$ scattering length\\ (-16.2 fm) is smaller than the 
accepted one ($-17.3\pm 0.4$ fm) although the error on the latter is 
perhaps optimistic. Our value does not depend on the CC potential
(removing it in the fit and the calculation of the ``strong''
calculation gives essentially the same value), nor does it depend
on the ranges $\mu_1$ and $\mu_2$ in Eqs.~(\ref{reidpot},\ref{ope1}) 
within limits. The difference
may very well come from the fact that we fit a different data
set than others. One sensitivity that was observed is to the
short-range Coulomb. For example, the use of a point Coulomb
potential would lead to a $pp$ scattering length of --16.7 fm.
In any case, the difference seems to represent a general shift 
since the $nn$ scattering length we obtain (~--18.0 \fmm) is also 
less negative than the measured value such that the breaking, often 
expressed as
\eq
|a_{nn}|-|a_{pp}|
\qe
is $1.8$ fm compared with $1.6\pm 0.6$ fm \cite{machmut}.

Up until this point we have said little about the spin of the diquark, 
it does not enter directly in the calculation. There is an implicit 
dependence, however. Since we assume that the odd quarks always couple 
to form the same correlation then, assuming an energy determined by the 
spin since the like quarks must couple to spin 1 the $ud$ quarks would 
need to be in a spin 1 state as well, the spin 0 and spin 1 diquarks 
would need to have the same interaction, or there could be some mixture 
of the two diquarks.

A number of papers have touched on this topic. Aside from those 
mentioned earlier, Close and Thomas \cite{close} studied quark 
distributions with an axial vector diquark heavier than the scalar 
diquark. Mineo et al. \cite{mineo2} consider a mixture and conclude 
that the axial vector diquark should have a weight of 2-10 \%. Clo\"et 
et al. \cite{bruno} find the inclusion of the axial vector diquark 
important for the nucleon electromagnetic form factors and the quark 
flavor distribution. Nagata and Hosaka \cite{nagata} find the axial 
vector diquark to be important for understanding the nucleon charge 
form factors. Finally, and perhaps most important, is the work of 
Clo\"et and Miller \cite{cloetmiller} (already mentioned in Section 
\ref{diquarkrev}) where the inclusion of the axial vector diquark leads 
to a possible understanding of the proton-spin puzzle. We see that 
recent work has indicated that the axial vector diquark may play a 
crucial role in the structure of the nucleon.

There would seem to be a fundamental difficulty with the assumption of 
only a scalar diquark for NN scattering. If the $ud$ quarks are assumed 
to form a scalar diquark then, in the $nn$ and $pp$ scattering the two 
``odd'' quarks must form a spin one object in the symmetric quark model 
(the axial diquark). One need not assume any clustering for this 
argument. Hence, the two nucleon system, which must be in a singlet 
state, cannot be formed. Thus, nucleons built up from a pure scalar 
diquark cannot generate the low-energy S-wave scattering. If the $ud$ 
pair always is a combination of scalar and axial vector diquarks then a 
coupling to total spin zero is possible for all three $NN$ pairs.

We finish with some general comments:

1) The CC model for isospin symmetry breaking depends only on the {\it 
density} of the quarks and not (directly at least) on their wave 
functions or spin. For this reason, obtaining the correct breaking is 
not a strong test of diquark models, although it provides a testing 
mechanism.

2) The non-zero-energy amplitude might also be calculated and may
give information about the rearrangement reaction. This might come
about through the study of the breaking in the effective range, $r_0$,
for example. A better understanding of the $nD$ breakup reaction
could provide crucial input in this regard.

3) Since models of the type we considered here are based only on the 
charges of the quarks, they are readily extensible to other sectors. 
The strange sector appears to be especially useful in this regard. 
Taking over results from the non-strange sector the CC concept is 
capable providing an understanding of the recent result of Gazda and 
Gal \cite{gazda} of a state-independent contribution to the breaking of 
the order of 200-300 keV. Of course other models may give a similar 
prediction but we know of no other at the present time.

One of us (WRG) acknowledges several very valuable conversations with 
B. F. Gibson. WRG also acknowledges the hospitality of the Laboratoire 
de Physique Nucl\'eaire et de Hautes \'Energies (LPNHE) where part of 
this work was done.

\begin{appendix}

\section{Analytic fits} In each case the parameters $C_1$ and $C_2$ 
are those in Eq. \ref{reidpot}, $M=2$\ \fmm. For reference, the 
original singlet S-wave Reid \cite{reid} soft-core potential had 
$C_1=-1650.6$ MeV and $C_2=6484.2$ MeV.

\cen{\large $\beta=12$ \fmm, p=0.58 fm$^3$}
$$
V_{CSB}(r)=C_{pp}(r)-V_C(r)-C_{nn}(r)
=\frac{0.35}{(1+e^{(r-.5)/1.2})}+0.07e^{-(r-1.75)^2/0.12\ }
$$
\eq
-0.019\ r \ e^{-10(r-3.5)^2}+0.04\ e^{-(r-1.8)^2/0.6}
\qe
\eq
C_{nn}(r)=0.21\ e^{-r^2/2.6}+0.011\ r\ e^{-(r-1.8)^2/0.6}
\qe
Parameters for $V_0(r)$: $C_1$=1757.87 MeV,\ $C_2$=7075.00 MeV

\cen{\large $\beta=8$ \fmm, p=1.0 fm$^3$}
\eq
V_{CSB}(r)=C_{pp}(r)-V_C(r)-C_{nn}(r)=\frac{0.19}{(1+e^{4(r-2.65)})}
+0.023\ e^{-12.5\ (r-1.78)^2}-0.019\ r\ e^{-10(r-0.95)^2}
\qe
\eq
C_{nn}(r)=0.226\ e^{-r^2/2.6}+0.011\ r\ e^{-(r-1.8)^2/0.6}
\qe

Parameters for $V_0(r)$:
 $C_1$=1792.88 MeV,\ $C_2$=7150.00 MeV

\cen{\large $\beta=4$ \fmm p=3.0 fm$^3$}
\eq
V_{CSB}(r)=C_{pp}(r)-V_C(r)-C_{nn}(r)=0.175\ 
e^{-(r-1.93)^2/0.75}-0.03\ e^{-1.2r}
\qe
\eq
C_{nn}(r)=0.17\ e^{-(r/1.2)^2}+0.05\ e^{(r-1.8)^2}
\qe
Parameters for $V_0(r)$:
 $C_1$=1809.57 MeV,\ $C_2$=7400.00 MeV
\
\end{appendix}

\end{document}